\documentclass[aps,prl,twocolumn,showpacs,floats]{revtex4}
\usepackage{amsmath}
\usepackage{tipa}
\usepackage{bbm}
\usepackage{txfonts}
\usepackage{graphicx}
\usepackage{dcolumn}
\usepackage{bm}
\usepackage{amssymb}
\usepackage{latexsym}
\usepackage{color}
\usepackage{subfigure}
\usepackage{ulem}

\newcommand{\veps}{\varepsilon}
\def\mbfe{\mathbf{e}}
\def\mbfB{\mathbf{B}}
\def\mbfF{\mathbf{F}}
\def\mbfn{\mathbf{n}}

\def\mbfr{\mathbf{r}}

\def\mbfa{\mathbf{a}}
\def\mbfg{\mathbf{g}}

\begin{document}

\title{Quantum Rotor Atoms in Light Beams with Orbital Angular Momentum: Highly Accurate Rotation Sensor}

\author{Igor Kuzmenko$^{1,2,4}$, Tetyana Kuzmenko$^{1,4}$, Y. B. Band$^{1,2,3,4}$}

\affiliation{
  $^1$Department of Physics,
  Ben-Gurion University of the Negev,
  Beer-Sheva 84105, Israel
  \\
  $^2$Department of Chemistry,
  Ben-Gurion University of the Negev,
  Beer-Sheva 84105, Israel
  \\
  $^3$Department of Electro-Optics,
  Ben-Gurion University of the Negev,
  Beer-Sheva 84105, Israel
  \\
  $^4$The Ilse Katz Center for Nano-Science,
  Ben-Gurion University of the Negev,
  Beer-Sheva 84105, Israel}

\begin{abstract}
Atoms trapped in a red detuned retro-reflected Laguerre-Gaussian beam undergo orbital motion within rings whose centers are on the axis of the laser beam.  We determine the wave functions, energies and degeneracies of such quantum rotors (QRs), and the microwave transitions between the energy levels are elucidated. We then show how such QR atoms can be used as high-accuracy rotation sensors when the rings are singly-occupied.
\end{abstract}

\pacs{32.80.Pj,71.70.Ej,73.22.Dj}

\maketitle

\begin{figure}%[htb]
\centering
\subfigure[]{\includegraphics[width=0.55\linewidth,angle=0] {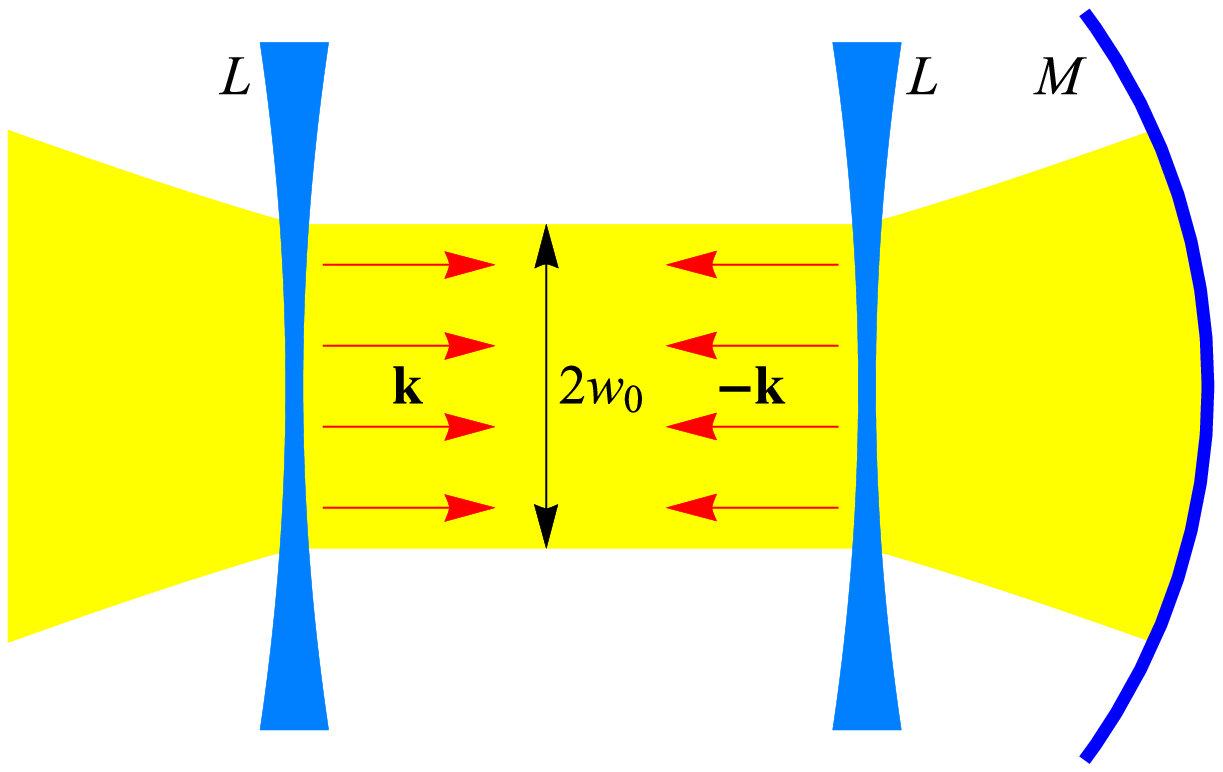}
  \label{Fig1a}}
\subfigure[]{\includegraphics[width=0.35\linewidth,angle=0] {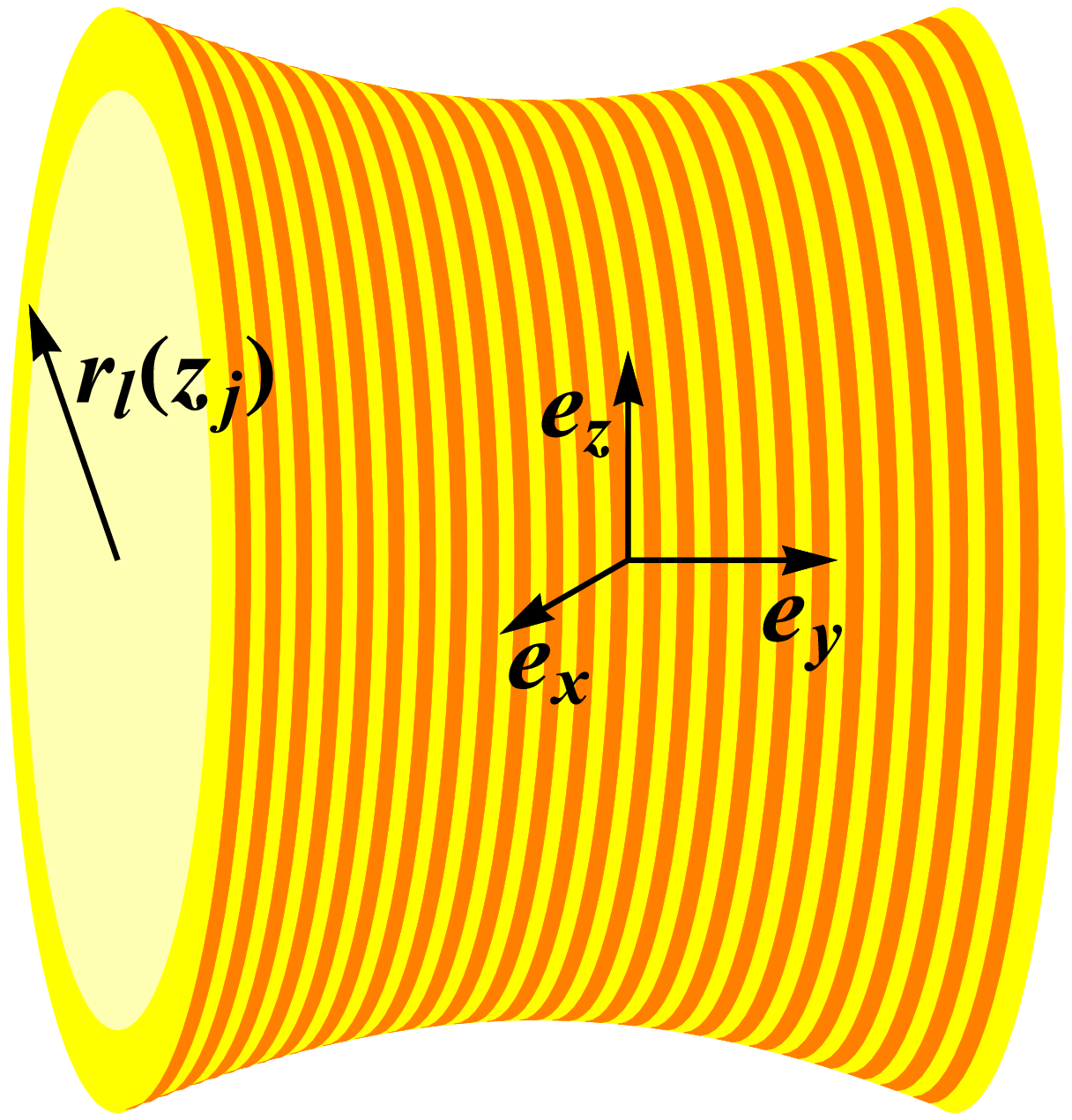}
  \label{Fig1b}}
\caption{ (a) Two lenses ($L$) refract the LG beam (yellow region).  
The mirror ($M$) reflects the beam, and two counter-propagating beams result. 
An almost uniform beam waist $w_0$ between the lenses, and a series of ring optical
potentials with zero amplitude at the center are stacked perpendicular to the beam axis.
The wave vectors of the incident and reflected beams are $\pm {\bf k}$.
(b) Blowup of the region between the two mirrors where the LG beam
forms the ring shaped potential wells.  Yellow and orange denote potential 
minima and maxima.  $\mbfe_x$, $\mbfe_y$ and $\mbfe_z$ are unit vectors parallel to 
the $x$, $y$ and $z$-axes, and $(r_l(z_j),z_j)$ are minima of the potential energy.}
\label{Fig1}
\end{figure}

{\it Introduction}:  We show that quantum rotor (QR) atoms (atoms whose motion is constrained to a circular ring) \cite{Kuzmenko_19} can be formed in light beams having orbital angular momentum and that they can be used as an extremely high accuracy rotation sensor.  QR atoms are trapped in a red-detuned linearly polarized retro-reflected Laguerre-Gaussian (LG) beam \cite{Goubau_61, Allen_92, Clifford_98, LG-beam-lens-2004, LG-beam-lens-1987}.  A line of singly-occupied rings filled with QR atoms can be easily formed (see Fig.~\ref{Fig1}) \cite{Viverit_04, Bloch_08, Gibbons_08}.  Single-occupation and negligible tunneling between rings are important to suppress deleterious spin exchange collisions between QR atoms for sensor applications.  The accuracy obtained here suggests that this can be the highest precision rotation sensor proposed so far in the literature.

A LG beam propagating along the $z$-axis with orbital angular momentum $l$ and polarization ${\bf e}_\alpha$ can be written in terms of a slowly varying envelope $u_{l,p}(r,\phi, z)$ of the electric field as
\begin{equation}  \label{eq:ELG}
{\bf E}_{\alpha,l,p}({\bf r},t) = u_{l,p}(r,\phi, z) \, e^{i (k z - k z_0 -\omega t)} \, {\bf e}_\alpha + {\mathrm{c.c.}},
\end{equation}
with field amplitude mode ${\rm{LG}}_p^l({\bf r}) \equiv u_{l,p}({\bf r})$ \cite{Goubau_61, Allen_92, Clifford_98},
\begin{equation}  \label{eq:LGM}
\begin{aligned}
u_{l,p}(r,\phi, z) = \sqrt{\frac{2p!}{\pi \left( {p + \left| l \right|!} \right)}} \frac{{\sqrt{P_0/c}}}{{w\left( z \right)}}{\left( {\frac{r\sqrt 2}{w\left( z \right)}} \right)^{\left| l \right|}}\exp \left( { - \frac{{{r^2}}}{{{w^2}\left( z \right)}}} \right)  \\ \times
L_p^{\left| l \right|}\left( {\frac{{2{r^2}}}{{{w^2}\left( z \right)}}} \right) \exp \left( { - \frac{{ik{r^2}z}}{{2\left( {{z^2} + z_R^2} \right)}}} \right)\exp \left( { - il\phi } \right) \\ \times
\exp \left[ {i\left( {2p + |l| + 1} \right){{\tan }^{ - 1}}\left( {\frac{z}{{{z_R}}}} \right)} \right] .
\qquad \qquad
\end{aligned}
\end{equation}
Here $z$ is the longitudinal distance from the beam waist
located at $z=0$, $P_0$ is the laser beam power,
$w_0$ is the beam waist at $z=0$, $R(z) = z(1+(z_R/z)^2)$ is
the radius of curvature of the beam wavefront,
$w(z) = w_0 (1+(z/z_R))^{1/2}$ is the radius at which
the beam intensity falls to $1/e$ of its axis value
at $z$, $z_R = \pi w_0^2/\lambda$ is the Rayleigh range for
the laser with wavelength $\lambda = 2\pi/k$ where
$k = \omega/c$ is the wavenumber, $0 < z_0 < \lambda/2$ is
a phase parameter, $L_p^{\left| l \right|}(x)$ is the associated
Laguerre polynomial, $\phi$ is the azimuthal angle, and
$\tan^{-1}\left(z/z_R\right)$ is the Gouy phase.
Figure~\ref{Fig1a} is a schematic diagram of a retro-reflected
LG beam propagating along the $z$-axis, and
Fig.~\ref{Fig1b} shows superposition of two counter-propagating
beams that form a standing wave along the $z$-axis.
The slowly varying envelope $u_{l,p}(r,\phi, z)$ of
the counter-propagating (cp) standing wave has the form
\begin{equation}  \label{eq:counter-prop-ELG}
{\bf E}^{\mathrm{cp}}_{\alpha,l,p}({\bf r},t) = u_{l,p}(r,\phi, z)  \,
{\bf e}_\alpha \, (e^{i (k z - k z_0 - \omega t)} + e^{i (- k z + k z_0 - \omega t)}).
\end{equation}
This standing wave configuration results in a series of ring shaped optical 
potentials [see the orange rings in Fig.~\ref{Fig1b}] stacked perpendicular to
the axis of the beams.  Since our interest is in trapping
atoms in the light beam, the light is red-detuned from atomic resonance, 
and atoms will be trapped in the ring shaped optical potentials that are singly
occupied so that spin relaxation collisions are suppressed \cite{Kuga_97}.

{\it QR Bound States in LG Rings}:  The QR Hamiltonian operator in cylindrical coordinates is
\begin{equation}  \label{Eq:Ham}
  H =
  -\frac{\hbar ^2}{2 M}
  \left(
       \frac{\partial ^2}{\partial r^2} +
       \frac{1}{r} \frac{\partial}{\partial r} +
       \frac{1}{r^2} \frac{\partial^2}{\partial \phi^2} +
       \frac{\partial^2}{\partial z^2}
  \right) +
  V\left(r,\phi,z\right),
\end{equation}
where the first term is the atom kinetic energy in polar
coordinates and $V(r,\phi,z)$ is the optical potential
resulting from the LG beams, which is calculated as
a second-order ac Stark shift \cite{SOI-EuroPhysJ-13}
and is given in terms of the ac polarizability $\alpha(\omega)$
by $V({\bf r}) = -\alpha(\omega) |{\bf E}^{\mathrm{cp}}_{\alpha,l,p}({\bf r},t)|^2$.
In the standing wave configuration in the nearly constant beam waist region,
the optical potential can be taken to be two-dimensional since
the width of the rings are very small.  For a ${\rm{LG}}_0^l$ mode
($p = 0$ and $l \neq 0$), the potential is independent of $\phi$,
\begin{eqnarray}
  V(r,z) = -V_0 ~ \cos^2\big( k ( z - z_0 ) \big)~
  \frac{\rho^{2 |l|}(z)}{\mathfrak{w}^2(z)} e^{-|l| (\rho^2(z) - 1)},
  \label{eq:V-isotrop-p=0}
\end{eqnarray}
where $\rho(z) = r/r_l(z)$ and $\mathfrak{w}(z) = w(z)/w_0$.
Potential (\ref{eq:V-isotrop-p=0}) has a minimum at
\begin{equation}
  z = z_j \equiv \frac{\pi}{k} j + z_0,
  \quad
  r = r_l(z_j) \equiv w(z_j) \sqrt{| l |/2},
  \label{eq:rL-zn-def}
\end{equation}
where $j$ is integer. The trapped atoms execute circular motion around
the $z$ axis, i.e., they are QRs.  $V(r_l(z_j),z_j)$ is given by
$V\big(r_l(z_j),z_j\big) = -V_0 ~ \big( w_0/w(z_j) \big)^{2}$.
For $z$ close to $z_j$,
\begin{equation}   \label{eq:V-r-z-harmonic}
  V(r,z) \approx V_l(r) + W_j(z),
\end{equation}
where
\begin{eqnarray}
  V_l(r) = V(r,z_j),
  \quad
  W_j(z) = \frac{V_0 k^2}{\mathfrak{w}^2(z_j)} ~ (z - z_j)^{2}.
  \label{eq:V-2D-xy-V-1D-z}
\end{eqnarray}
$W_j(z)$ is a harmonic potential in $z - z_j$.
The corresponding harmonic frequency and length are
\begin{eqnarray}
  \omega_z(z_j) = {\frac{2}{\mathfrak{w}(z_j)} 
  \frac{\sqrt{{\mathcal{E}}_{0} V_0}}{\hbar},
  \quad
  b_z(z_j) =
  \frac{\sqrt{\mathfrak{w}(z_j)}}{k}}
  \bigg(\frac{{\mathcal{E}}_{0}}{V_0}\bigg)^{1/4},
  \label{eq:Omega-z-b-z}
\end{eqnarray}
where ${\mathcal{E}}_{0} = \hbar^2 k^2/( 2 M )$ is the recoil energy.

%------------------------------ energies inertial ------------------------------
\begin{figure}%[htb]
\centering
  \includegraphics[width=0.8\linewidth,angle=0]
   {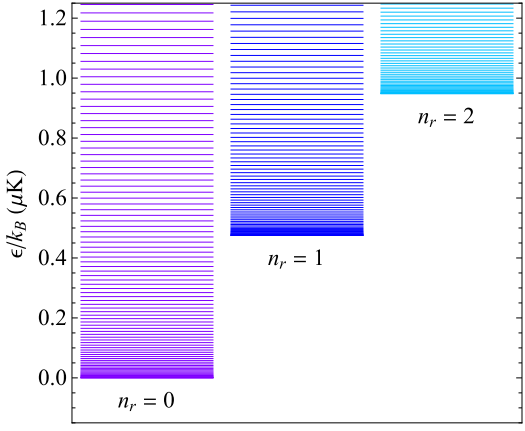}
\caption{Energies $\epsilon(n_z,n_r,m_{\ell})$ of the QR
(relative to the QR ground state energy) trapped in an LG beam with $w_0 = 10~\mu$m, 
$l = 5$, $p = 0$, and $V_0 = 10 ~ {\mathcal{E}}_{0}$, where ${\mathcal{E}}_{0}$ is the recoil energy.
This gives $r_l \equiv r_l(z_j)$ with $j=0$ equal to $15.81~\mu$m.  Energies with
$n_r = 0$ are shown as purple lines, $n_r = 1$ as blue lines and $n_r = 2$ as sky-blue lines. 
Quantum states with $n_z \geq 1$ have very high energies and fall out of the scale of the figure,
hence only $n_z = 0$ are shown.}
\label{Fig:energies}
\end{figure}

The optical potential (\ref{eq:V-isotrop-p=0}) is invariant
with respect to rotations about the $z$ axis,
therefore the quantum states of
the QR in the harmonic approximation (\ref{eq:V-r-z-harmonic})
are parametrized by radial and vertical quantum numbers
$n_r$ and $n_z$ describing radial and $z$ motion ($n_r,n_z = 0,1,2,\ldots$),
the orbital momentum quantum number $m_{\ell}$ and projection $m_F$ of the hyperfine 
orbital momentum ${\bf F}$ on the $z$ axis.  The ground state has 
$n_z = n_r = m_{\ell} = 0$, and is $2F+1$ fold degenerate.  
Orbitally excited states with $m_{\ell} \neq 0$ are $2(2F+1)$ fold degenerate, 
and have angular momentum $\pm m_{\ell}$. 
Radial and  vertical excitations have $n_r \neq 0$ and $n_z \neq 0$ respectively.  For 
simplicity, in this paragraph, let us consider an atom trapped at the $z_0$ site 
(i.e., near $z_j$ with $j = 0$).
The QR wave functions and eigen-energies satisy the Schr\"odinger equation,
\begin{equation}    \label{eq:Schrodinger}
  \Big[-\frac{\hbar^2}{2 M}~ \nabla^2 + V(r,z) - \epsilon_{\bf n} \Big]
  \Psi_{\bf n}(\mbfr) = 0 ,
\end{equation}
where $\mbfn = (n_z,n_r,m_{\ell})$.  The wave function can be written in cylindrical 
coordinates $\mbfr = (r,\phi,z)$ as
\begin{equation}   \label{WF-rad-ang-spin}
  \Psi_{\mbfn}(\mbfr) =
  \frac{1}{\sqrt{2 \pi}}~
  \eta_{n_z}(z)
  \psi_{n_r,m_{\ell}}(r) e^{i m \phi},
\end{equation}
where $\eta_{n_z}(z)$ and $\psi_{n_r,m_{\ell}}(r)$
satisfy the equations,
\begin{equation}  \label{eq:Schrodinger-z}
  \Big[ -\frac{\hbar^2}{2 M}~ \frac{d^2}{dr^2} +
  W_0(z) - \epsilon_{z}(n_z) \Big] \eta_{n_z}(z) = 0 \, ,
\end{equation}  
\begin{equation}  \label{eq:Schrodinger-radial}
  \Big[-\frac{\hbar^2}{2 M r}~\frac{d}{dr}\bigg( r~ \frac{d}{dr} \bigg) +
  V_l(r) + m_{\ell}^{2}~C(r) -\epsilon_{r}(n_r,m_{\ell}) \Big]~\psi_{n_r,m_{\ell}}(r) = 0 .
\end{equation}
$V_l(r)$ and $W_j(z)$ are given by Eq.~(\ref{eq:V-2D-xy-V-1D-z}), and 
$C(r) = \frac{\hbar^2}{2 M r^2}$ is the rotational energy of the QR around the $z$ axis.  
The eigen-energy of the trapped atom is
\begin{eqnarray}
  \epsilon(\mbfn) &=&\epsilon_{z}(n_z) + \epsilon_{r}(n_r,m_{\ell}),
  \label{eq:energoes-nz-nr-m}
\end{eqnarray}
where $r_l \equiv r_l(z_j)$ with $j=0$.
The QR vertical, radial and orbital excitation energies are
\begin{eqnarray*}
  \veps_z &=&
  \epsilon_z(1) -
  \epsilon_z(0)
  ~\approx~\hbar \omega_z,
  \\
  \veps_{r} &=&
  \epsilon_r(1,0) -
  \epsilon_r(0,0),
  \\
  \veps_{\ell} &=&
  \epsilon_r(0,1) -
  \epsilon_r(0,0)
  ~\approx~ C(r_l).
\end{eqnarray*}
We assume they satisfy the inequalities,
\begin{equation}
  \veps_z ~\gg~
  \veps_r ~\gg~
  \veps_{\ell},
  \label{inequalities:energies}
\end{equation}
i.e., the orbital excitations are the lowest-energy excitations,
whereas the radial and longitudinal excitations have relatively
high energies.
The energies $\epsilon(n_z = 0,n_r,m_{\ell})$, Eq.~(\ref{eq:energoes-nz-nr-m}),
are shown in Fig.~\ref{Fig:energies}.  For $l = 5$, $w_0 = 10~\mu$m,
$r_l = 15.81~\mu$m and $V_0 = 10 ~ {\mathcal{E}}_{0} = k_B \times 35.36~ \mu$K,
the excitation energies are $C(r_l) = k_B \times 0.1613$~nK,
$\hbar \omega_r = k_B \times 0.4776~\mu$K and
$\hbar \omega_z = k_B \times 22.36~\mu$K,
so the inequalities (\ref{inequalities:energies}) are valid.
Quantum states with $n_z \geq 1$ have high energies
and are not shown in Fig.~\ref{Fig:energies}.

%---------------- QR-Rabi-opt ---------------------------
\begin{figure}[htb]
%H=6.28, L=14.65
\centering
  \subfigure[]
  {\includegraphics[width = 70 mm,angle=0]
   {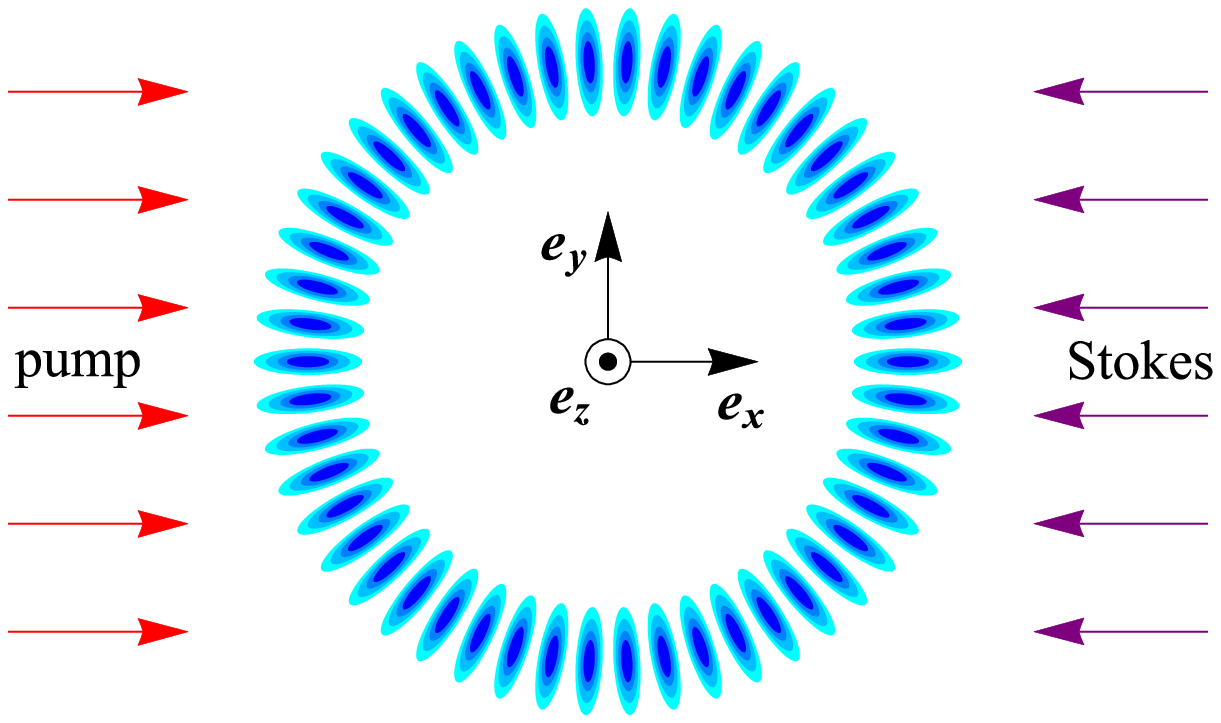}
   \label{Fig-QR-three-pulses}}
  \subfigure[]
  {\includegraphics[width = 70 mm,angle=0]
   {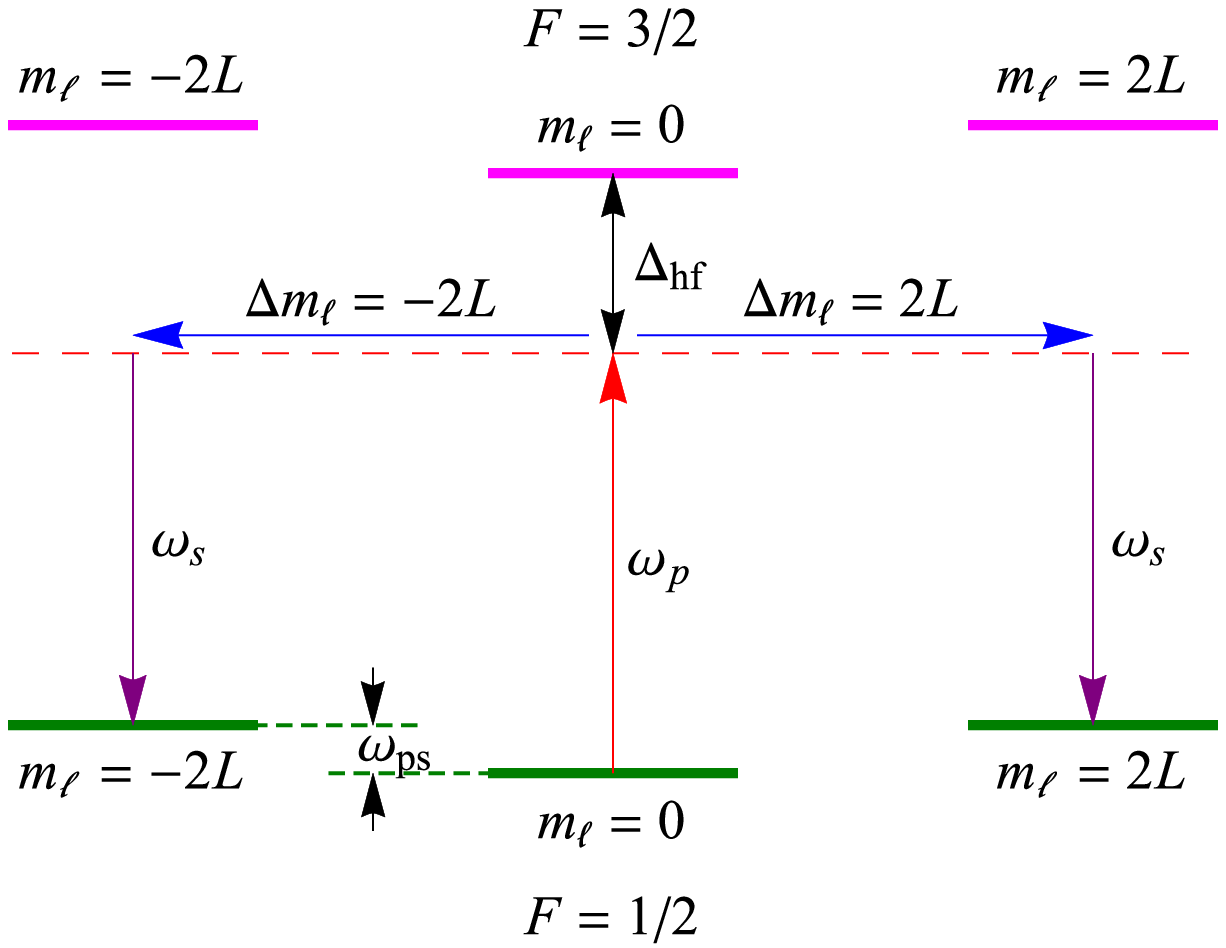}
   \label{Fig-Rabi-oscillations}}
 \caption{(a) Pump and Stokes radio-frequency pulses (red and purple arrows)
   and optical rotational counter-propagating LG rotational-kick-pulses
   with $L = 25$ along the $z$ axis. 
   The blue regions indicate the depth of the rotational-kick optical pulse potential.
   $\mbfe_x$, $\mbfe_y$ and $\mbfe_z$ are unit vectors along the $x$-, $y$- and $z$-axes.
   (b) Quantum transitions due to the pump, Stokes and rotational-kick
   pulses (red, purple and blue arrows). The frequencies of the pump and Stokes 
   pulses are $\omega_p$ and $\omega_s$, and their difference $\omega_{ps} = \omega_p 
   - \omega_s$ is equal to the transition frequency $\omega_{2L,0}$.
   Detuning of $\omega_p$ from the resonant frequency
   of the ${}^{2}$S$_{1/2}(F=1/2)$ to ${}^{2}$S$_{1/2}(F=3/2)$
   quantum transition is $\Delta_{\mathrm{hf}}$.}
\label{Fig-QR-Rabi-radio-opt}
\end{figure}

{\it Rabi Oscillation Method with Raman Pulses}:
In order to measure the excitation energies of the QR atoms with
quantum numbers $n_r$, $m_{\ell}$ and $m_F$, we propose
to subject the QRs to three pulses, as shown in Fig.~\ref{Fig-QR-three-pulses}:
pump and Stokes radio-frequency pulses with frequencies
$\omega_p \gtrapprox \omega_s$ that are detuned from the quantum
transition between the hyperfine states of the ground state, and
a LG pulse with frequency $\omega_e$
far-detuned from the resonant frequency of
the $^2$S$_{1/2} \rightarrow {}^2$P$_{3/2}$ electronic transition.
The radio-frequency pump and Stokes light cannot change
the orbital quantum number $m_{\ell}$.
Therefore we use a LG pulse that allows
quantum transitions $|m_{\ell}\rangle \to |m'_{\ell}\rangle$
between quantum states with different orbital quantum
numbers $m_{\ell}$ and $m'_{\ell}$ [see Fig.~\ref{Fig-Rabi-oscillations}].
The pump and Stokes pulses propagate parallel and anti-parallel to 
the $x$-axis and are linearly polarized with magnetic
field along the $z$ axis (see the supplemental materials (SM) \cite{SM}).
The pump and Stokes magnetic fields are given by
$\mbfB_{\mu}(\mbfr,t) = \mbfB_{\mu}^{(0)} ~ \cos(k_{\mu} x - i \omega_{\mu} t)%
\Theta(t) \Theta(\tau - t) \approx \mbfB_{\mu}^{(0)} \cos(\omega_{\mu} t) 
\Theta(t) \Theta(\tau - t)$, where $\mu = p,s$ (pump and Stokes beams) and
we have assumed square pulses, hence the presence of the $\Theta$ functions.
Here we take into account that the wavelengths $\lambda_{\mu}$
are much longer than the radius $r_l$ of the QR, hence
$\mbfB_{\mu}$ depends on $t$, but not on $\mbfr$, and therefore
these pulses do not possess Raman transitions between the quantum states
$|n_r,m_{\ell}\rangle$ and $|n'_r,m'_{\ell}\rangle$ and we can approximate
$\mbfB_{\mu}(\mbfr,t)$ by $\mbfB_{\mu}(t) \equiv \mbfB_{\mu}(0,t)$.
Hence, the dipole magnetic interaction between the QR and
the radio waves, $H_{\mu} = -g_F \mu_B \mbfF \cdot \mbfB_{\mu}(t)$,
does not depend on $\mbfr$ (the position of the atom), and therefore
$\langle n_r,m_{\ell} | H_{\mu} | n'_r,m'_{\ell} \rangle  \propto %
\delta_{n_r,n'_r} \delta_{m_{\ell},m'_{\ell}}$.  For Raman transitions 
$|n_r,m_{\ell}\rangle \to |n_r,m'_{\ell}\rangle$ with $m_{\ell} \ne m'_{\ell}$,
an optical square pulse which breaks the cylindrical symmetry of the QR is required.
The electric field of the optical pulse is ${\bf E}_e(\mbfr,t) = \frac{1}{2} 
(u_{L,0}(\mbfr) + u_{-L,0}(\mbfr))~ \cos(k_e z) ~ \Theta(t) \Theta(\tau - t) 
~ e^{-i \omega_e t}~ \mbfe_x + {\mathrm{c.c.}}$, where $k_e = \omega_e/c$ 
is the wavenumber of the optical pulse and
$u_{L,0}(\mbfr)$ is given by Eq.~(\ref{eq:LGM}).
We choose the waist radii of the LG pulse $w_e$ and the LG beam waist $w_0$,
and their orbital angular momentums $L$ and $l$ in such a way that
$w_e \sqrt{|L|/2} = w_0 \sqrt{|l|/2}$.

The stimulated Raman process corresponding to absorption of
a pump photon and stimulated emission of a Stokes photon gives
rise to excitation of the QR.  This scattering is described by
the time-dependent Hamiltonian with matrix elements given by
\begin{equation}
  {\mathcal{H}}_{n_r,m_{\ell};n'_r,m'_{\ell}}(t) =
  -2 {\mathcal{V}}
  \cos(\omega_{ps} t)\Theta(t) \Theta(\tau - t)
  \delta_{n_r,n'_r} \sum_{m = \pm 2 L} \delta_{m_{\ell},m'_{\ell} + m},
  \label{eq:H-Rabi}
\end{equation}
where $\omega_{ps} = \omega_p - \omega_s$ is the difference of
frequencies of the pump and Stokes pulses, and
\begin{equation}   \label{eq:V}
  {\mathcal{V}} =
  \frac{ {\mathcal{V}}_{e} {\mathcal{V}}_{b} }{\hbar \Delta_{\mathrm{hf}}},
  \ \
  {\mathcal{V}}_b =
  \frac{g^2 \mu_{B}^{2} B_{p}^{(0)} B_{s}^{(0)}}{3 \hbar \Delta_{\mathrm{hf}}},
  \ \
  {\mathcal{V}}_e =
  \frac{4 \alpha(\omega_e)}{\pi L!}~
  \frac{P_e L^L e^{-L}}{w_{e}^{2} c},
\end{equation}
$P_e$ and $w_e$ are the power and the beam waist of the optical pulse,
$B_{p}^{(0)}$ and $B_{s}^{(0)}$ are the magnetic field strengths
of the pump and Stokes radio-frequency pulses.
The subscript $e$ symbolizes the electric dipole interaction of the atom with
the optical pulse, and the subscript $b$ symbolizes the magnetic dipole interaction
of the atoms with the pump and Stokes radio-frequency pulses.
The detuning $\Delta_{\mathrm{hf}}$ of the pump pulse frequency
from the ${}^{2}$S$_{1/2}(F = 1/2) \to {}^{2}$S$_{1/2}(F = 1/2)$
quantum transition frequency greatly exceeds
the frequency $\omega_{2L,0}$ of the $|0\rangle \to |2L\rangle$ 
quantum transition,
\begin{equation}    \label{inequalities-Delta_e-gg-Delta_hf-gg-omega_2L}
  \Delta_{\mathrm{hf}} ~\gg~ \omega_{2L,0},
\end{equation}
hence we can assume that $\omega_p$ and $\omega_s$
have the same detuning $\Delta_{\mathrm{hf}}$ from resonance.
Details of the Raman scattering of the pump and Stokes light, the Rabi 
oscillations with Raman pulses used to measure the energy
differences of the QR states are presented Ref.~\cite{Kuzmenko_19} and also 
in the SM \cite{SM}.

Consider the QR initially in the ground state $|0\rangle$ subjected
to Raman pulses of duration $\tau$ and generalized Rabi frequency
$\Omega_R = 2 \sqrt{2} {\mathcal{V}}/\hbar$, such that
$\Omega_R \tau \approx \pi$.  In the 3-level rotating wave 
approximation \cite{rotating-wave-approx-PRL-2007}, 
the temporal evolution of a single QR wave function is given by
$| \Psi(\tau) \rangle = e^{-i {\mathcal{H}}_{R} \tau/\hbar} | 0 \rangle$,
where ${\mathcal{H}}_{R}$ is the Hamiltonian (see SM \cite{SM}),
\begin{eqnarray}
  {\mathcal{H}}_{R} &=&
  \hbar
  \left(
    \begin{array}{ccc}
      0 & \Omega_R\sqrt{2}/4 & \Omega_R\sqrt{2}/4
      \\
      \Omega_R\sqrt{2}/4 & -\delta & 0
      \\
      \Omega_R\sqrt{2}/4 & 0 & -\delta
    \end{array}
  \right),
  \label{eq:H-Ramsey}
\end{eqnarray}
$\delta = \omega_{ps} - \omega_{2L,0}$, and the basis vectors are,
$$
  | 0 \rangle = \left( \begin{array}{c} 1 \\ 0 \\ 0 \end{array} \right),
  \quad
  | 2 L \rangle = \left( \begin{array}{c} 0 \\ 1 \\ 0 \end{array} \right),
  \quad
  | -2 L \rangle = \left( \begin{array}{c} 0 \\ 0 \\ 1 \end{array} \right).
$$
The probability of finding the quantum rotor in the final state
$|f\rangle = 2^{-1/2}( | 2L \rangle + | -2L \rangle )$ is
\begin{equation}    \label{eq:S0-single-QR}
  P_0(\delta,\Omega_R) =
  \big| \langle f | \Psi(\tau) \rangle
  \big|^{2} = \frac{\Omega_{R}^{2}}{\Omega_{R}^{2} + \delta^2}
  \sin^2 \bigg( \frac{\tau}{2} \sqrt{\Omega_{R}^{2} + \delta^2} \bigg).
\end{equation}
$P_0(\delta,\Omega_R)$ has a peak, $P_0(0,\Omega_R)=1$
at $\delta=0$. The peak width at half maximum is $1.597 ~ \Omega_R$.

When we have $N = 2 j_{\max} + 1$ atoms singly occupying
the sites with $|j| \leq j_{\max}$, the detuning of $\omega_{ps}$
from the resonant frequency of the $j$-th quantum rotor is
$\delta_j = \delta + 4 L^2 j^2 \big( \omega_0(r_l(0)) - \omega_0(r_l(z_j)) \big)$,
where $\delta$ is the detuning of $\omega_{ps}$ from the resonant
frequency of the $j = 0$ QR, $\omega_0(r_l(z)) = \hbar/(2 M r_{l}^{2}(z_j))$ and
$r_l(z_j)$ is the radius of the classical circular trajectory.
As a result, the peak in the probability to find a QR
in the final state is shifted and broadened.
The probability to find a QR in the final state is
\begin{eqnarray}
  P(\delta,\Omega_R) &=&
  \frac{1}{2 j_{\max} + 1}
  \sum_{j = -j_{\max}}^{j_{\max}}
  P_0\big( \delta_j,\Omega_R \big).
  \label{eq:S-omega}
\end{eqnarray}

%---------------- S-omega ---------------------------
\begin{figure}[htb]
%H=6.28, L=14.65
\centering
  \includegraphics[width=0.9\linewidth,angle=0]
   {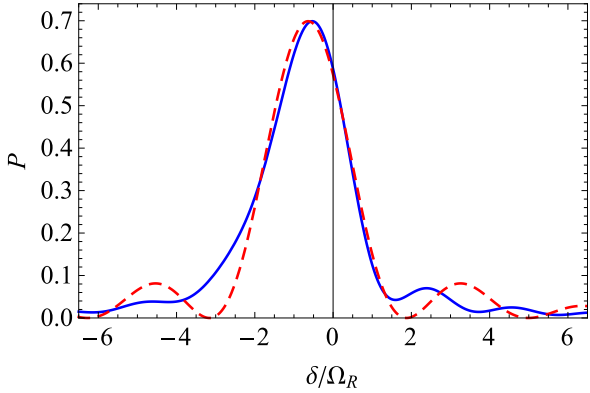}
\caption{The probability $P$ in Eq.~(\ref{eq:S-omega}) for absorption of
   a pump photon and stimulated reemission of a Stokes photon with
   QRs quantum transition from state $|m_{\ell} = 0\rangle$
   to $|m'_{\ell} = 50\rangle$ as a function of the difference
   $\omega_{ps} = \omega_p - \omega_s$ of the pump and Stokes
   frequencies $\omega_p$ and $\omega_s$ (solid blue curve).
   The dashed red curve shows the fit of $P(\delta,\Omega_R)$ 
   obtained as explained in the text.}
\label{Fig-S-omega}
\end{figure}

Figure~\ref{Fig-S-omega} shows $P(\delta,\Omega_R)$ of Eq.~(\ref{eq:S-omega})
plotted as a function of $\delta$, for
$\omega_0 = 21.13$~s$^{-1}$, which corresponds to a LG beam
with $w_0 = 10 ~ \mu$m and $l = 25$,
Rabi frequency $\Omega_R = 3.142$~s$^{-1}$, pulse duration 
$\tau \approx \pi/\Omega_R = 1$~s,
$j_{\max} = 80$, $m_{\ell} = 0$ and $m'_{\ell} = 50$.
The solid blue curve shows that $P(\delta,\Omega_R)$ has a peak,
$P_{\max} = P(\delta_{\max},\Omega_R) = 0.6989$, at
$\delta_{\max} = -0.5374 \Omega_R$.
The dashed red curve shows the fitting of $P(\delta,\Omega_R)$
by the function $P(\delta,\Omega_R) \approx {\mathcal{A}} \,
P_0(\delta-\delta_0,\tilde\Omega_R)$ with ${\mathcal{A}} = 0.67987$ 
and $P_0$ is given by Eq.~(\ref{eq:S0-single-QR}) with $\delta_0 = -0.64002 ~ \Omega_R$,
$\tilde\Omega_R = 1.4865 ~ \Omega_R$.  Note that
$\delta_{\max} \neq \delta_0$. This is partly because the function
$P(\delta,\Omega_R)$ is not symmetric with respect to
the inversion $\delta - \delta_{\max} \to -( \delta - \delta_{\max})$,
whereas the function $P_0(\delta,\Omega_R) = P_0(-\delta,\Omega_R)$
is symmetric.

We now show that ground-state QRs in an LG beam can serve as extremely accurate rotation sensors.

{\it Rotation Sensor}:
Consider a QR in a non-inertial frame rotating with angular
velocity $\boldsymbol\Omega = \Omega ~ \mbfe_z$ (see Fig.~2 in the SM \cite{SM}).
The additional term needed in the Hamiltonian is
\begin{eqnarray}
  H_{\Omega} = \hbar \Omega \ell_z,
\label{eq:H-Omega}
\end{eqnarray}
where $\ell_z = -i \partial_{\phi}$ is the orbital momentum operator.
The Hamiltonian (\ref{eq:H-Omega}) lifts the symmetry under the transformation 
$(x, y, z) \to (-x, y, z)$ 
but not the rotational symmetry about the $z$-axis.  As a result, $m_{\ell}$,
the eigenvalue of $\ell_z$, is a good quantum number, and the energy
levels $\epsilon_{m_{\ell}}$ of the rotational motion become
$\epsilon_{m_{\ell}}(\Omega) = \epsilon_{m_{\ell}} + \hbar \Omega \, m_{\ell}$.
Hereafter we use the inequalities (\ref{inequalities:energies})
and restrict ourselves by considering the quantum states with
$n_z = n_r = 0$.
Moreover, using the inequalities  (\ref{inequalities:energies}),
we approximate the rotational energies as $\epsilon_{m_{\ell}} = m_{\ell}^{2} C(r_l)$,
where $r_l$ is given by Eq.~(\ref{eq:rL-zn-def}).

\begin{figure}[htb]
%H=6.28, L=14.65
\centering
  \includegraphics[width=0.8\linewidth,angle=0]
   {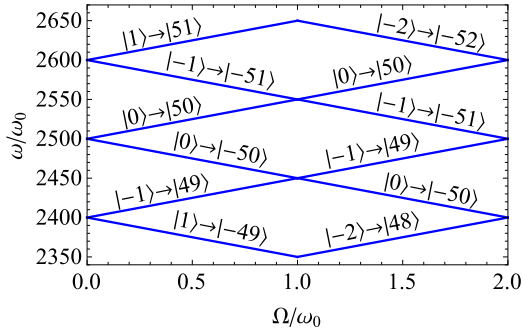}
 \caption{The frequencies (\ref{eq:Omega-vs-rotations}) of
   the quantum transitions $|m_{\ell}\rangle \to |m_{\ell} \pm 50\rangle$ versus the magnitude 
   of the rotational velocity $\Omega$.  $\omega_0$ is given by Eq.~(\ref{eq:a0-Omega0-def}).}
   \label{Fig-rotation-Omega}
\end{figure}

Let us consider the frequencies of the quantum transitions between
the quantum states $|m_{\ell}\rangle$ and $|m'_{\ell}\rangle$ with
$m_{\ell} = 0, \pm 1$ and $m' _{\ell}= m_{\ell} \pm 2 L$.
We have six spectral lines with frequencies $\omega_{m_{\ell}, m_{\ell} \pm 2 L}(\Omega)$ 
which are convenient to order as follows:  
$\omega_{0, \pm 2 L}(\Omega)$, $\omega_{\pm 1, \pm (1 + 2 L)}(\Omega)$,
and $\omega_{\pm 1, \pm (1 - 2 L)}(\Omega)$.  The frequencies of the quantum 
transitions between
$| \zeta m_{\ell} \rangle$ and $| \zeta m_{\ell} + 2 \zeta L \rangle$ are
\begin{equation}   \label{eq:Omega-vs-rotations}
  \omega_{m_{\ell},  m_{\ell} + 2 \zeta L}(\Omega) = 
  4 L \big(L + m_{\ell}\big)
  \omega_0 + 2 \zeta L \Omega.
\end{equation}
where $\zeta = \pm 1$, and the  rotational frequency is
\begin{eqnarray}  \label{eq:a0-Omega0-def} 
\omega_0 = C(r_l)/\hbar.
\end{eqnarray}
Hence, when $\Omega = 0$, $\omega_{m_{\ell}, m_{\ell} + 2 L}(0) = \omega_{-m_{\ell}, -(m_{\ell} + 2 L)}(0)$.
One can measure the three spectral lines $\omega_{0, 2 L}(0) = \omega_{0, -2 L}(0)$,
$\omega_{1, 1 - 2 L}(0) = \omega_{-1, -1 + 2 L}(0)$, and
$\omega_{1, 1 + 2 L}(0) = \omega_{0, -1 - 2 L}(0)$.
When $\Omega \neq 0$, the degeneracy of the spectral lines is lifted
and the splitting is
\begin{eqnarray}
  \Delta \omega_{m_{\ell},m_{\ell} + 2 L} \equiv
  \omega_{m_{\ell},m_{\ell} + 2 L}(\Omega) -
  \omega_{-m_{\ell},-m_{\ell} - 2 L}(\Omega)
  =
  4 L \Omega.
  \label{eq:Delta-omega-vs-Omega}
\end{eqnarray}
Eq.~(\ref{eq:Delta-omega-vs-Omega}) shows that measuring the splitting of
the spectral lines (\ref{eq:Omega-vs-rotations}), can be used to determine
$\Omega$.
Figure~\ref{Fig-rotation-Omega} shows the transition frequencies of
the $|m_{\ell}\rangle \to |m_{\ell} \pm 50\rangle$ transitions as functions 
of $\Omega$.  The frequency splitting (\ref{eq:Delta-omega-vs-Omega})
due to $\Omega$ distinguishes between clockwise and counter-clockwise rotations.
All the spectral lines have the same splittings. Hence, the frequencies 
satisfy the periodic condition,
$$
  \omega_{m_{\ell} + m_{\Omega},m_{\ell} + m_{\Omega} + 2 L}
  \big( \Omega - 2 m_{\Omega} \omega_0 \big)
  ~=~
  \omega_{m_{\ell},m_{\ell} + 2 L} \big( \Omega \big),
$$
where $m_{\Omega}$ is integer.

%---------- rotations-Omega ----------

{\it Rotation measurement accuracy estimate}:
Note that Eq.~(\ref{eq:Delta-omega-vs-Omega}) does not contain any
information regarding the optical potential, the laser frequency or the intensity.
Therefore the uncertainty of $\Omega$, $\delta \Omega$, is determined solely by uncertainty
$\delta \omega$ of the pump and Stokes frequencies,
\begin{eqnarray}
  \delta \Omega &=&
  \frac{1}{\sqrt{N}} \,
  \frac{\delta \omega}{4 L}.
  \label{eq:delta-Omega}
\end{eqnarray}
Here $N = 2 j_{\max} + 1$ is the number of atoms singly occupying the
sites with $|j| \leq j_{\max}$, $\delta \omega = \delta\omega_p + \delta\omega_s$, 
and $\delta\omega_p$ and $\delta\omega_s$ are the uncertainties
of the pump and Stokes frequencies. For ${}^{6}$Li atoms, we take
$\omega_p ~\gtrapprox~ \omega_s ~\approx~ 1.43 \times 10^9~{\text{s}}^{-1}$.
$\delta\omega$ can be estimated as
$\delta \omega = 2 \times 10^{-18} \omega_p \approx 2.86 \times 10^{-9}~{\text{s}}^{-1}$.
From Eq.~(\ref{eq:delta-Omega}) we see that the larger $L$,
the smaller is $\delta \Omega$; when $L = 25$
and $j_{\max} = 80$, $\delta \Omega = 2.25 \times 10^{-12}$~s$^{-1}$.

In order to measure angular velocity when gravity $\mbfg$ is present, place the 
QRs in the plane perpendicular to $\mbfg$ (let us call this the $x$-$y$ plane),
in order to measure the component of the angular velocity in the $z$ direction.
If an additional acceleration $\mbfa$ in the $x$-$y$ plane is present, there is
an additional splitting of the QR ground state degeneracy due 
to the acceleration, and the frequency splitting 
in Eq.~(\ref{eq:Delta-omega-vs-Omega}) becomes dependent on $m_{\ell}$.
Hence, turn the plane of the QRs to be perpendicular to 
${\mathbf{g}}' = \mbfg - \mbfa$
so that the frequency splitting (\ref{eq:Delta-omega-vs-Omega})
is independent on $m_{\ell}$, and obtain the energy splitting of the  
levels caused just by $\Omega' = \boldsymbol\Omega \cdot \mbfe'_z$,
where $\mbfe'_z$ is the unit vector along $-\mbfg'$.
For details see the SM \cite{SM}.

{\it Summary and Conclusion}:
Cold atoms trapped in a Laguerre-Gauss optical potential (\ref{eq:V-isotrop-p=0}) 
are confined to circular rings (donuts) of radius $r_l$ with centers on the axis of
the Laguerre-Gauss beam.  Rings with one atom per site (to suppress spin-exchange 
collisions) can be used as high-accuracy rotation sensors.
When $r_l = 15.81~\mu$m, the accuracy obtained with $N=161$ atoms singly occupying
the sites with $|j| \leq 80$ is $\delta \Omega = 2.25 \times 10^{-12}$~s$^{-1}$.
This is better than the accuracy $\delta \Omega = 6.4 \times 10^{-10}$~s$^{-1}$
reported in Ref.~\cite{Kuzmenko_19}.  Moreover, the rotation sensor accuracy is
much better than $\delta \Omega_{\mathrm{NIST}} = 3 \times 10^{-8}$~s$^{-1}$
reported in Ref.~\cite{rotation-sensor-NIST-2016}.

The rings between the lenses in Fig.~\ref{Fig1} have slightly
different radii [see Eq.~(\ref{eq:rL-zn-def})].
As a result, different transition frequencies are obtained from of the
$m_{\ell} \to m'_{\ell}$ transition for different $z_j$:
$\omega_{m_{\ell},m'_{\ell}}(z_j) = \omega_{m_{\ell},m'_{\ell}}(0) C(r_l(z_j))/C(r_l(0))$.
Hence, the width of the Rabi oscillation is broadened.
For example, when $\omega_0 = 21.13$~s$^{-1}$, $\Omega_R = 3.142$~s$^{-1}$,
and there are $N = 161$ singly occupied sites with $|j| \leq 80$, the resulting 
effective width of the Rabi oscillation peak is $\tilde\Omega_R = 1.4865 ~ \Omega_R$
instead of $\Omega_R$ for a single QR (see Fig.~\ref{Fig-S-omega}).

\begin{acknowledgments}
This work was supported in part by a grant from the DFG through the DIP program (FO703/2-1).
\end{acknowledgments}

%%%%%%%%%%%%%%%%%%%%%%%%%%%%%%

\end{document}

\title{Supplemental Material for ``Quantum Rotor Atoms
  in Light Beams with Orbital Angular Momentum:
  Highly Accurate Rotation and Acceleration Sensing''}

\author{Igor Kuzmenko$^{1,4}$, Tetyana Kuzmenko$^{1,4}$, Y. B. Band$^{1,2,3,4}$}

\affiliation{
  $^1$Department of Physics,
  Ben-Gurion University of the Negev,
  Beer-Sheva 84105, Israel
  \\
  $^2$Department of Chemistry,
  Ben-Gurion University of the Negev,
  Beer-Sheva 84105, Israel
  \\
  $^3$Department of Electro-Optics,
  Ben-Gurion University of the Negev,
  Beer-Sheva 84105, Israel
  \\
  $^4$The Ilse Katz Center for Nano-Science,
  Ben-Gurion University of the Negev,
  Beer-Sheva 84105, Israel}

\maketitle

Here we expand the discussion of the main text \cite{main}
as follows.  Section \ref{sec:H-Rabi} contains details regarding
the Rabi oscillation method used to probe the energy
levels of quantum rotors (QR) trapped in Laguerre-Gaussian (LG) beams
to induce stimulated Raman transitions with radio-frequency 
pulses and optical frequency Laguerre-Gaussian pulses.
Section~\ref{sec:accuracy} augments the analysis of the accuracy of the 
rotation sensor which uses QR atoms in LG beams, as proposed in \cite{main}.
Section~\ref{sec:acceleration} considers the discrimination against 
the effects of in-plane acceleration on the rotation sensor.

%%%%%%%%%%%%%%%%%%%%%%%%%%%%%%
\section{Rabi Oscillations of Quantum Rotor States in LG Beams}
  \label{sec:H-Rabi}

Here derive the time-dependent Hamiltonian in Eq.~(16) of the main text \cite{main}
and treat the stimulated Raman process that gives rise to absorption of
a pump photons and stimulated emission of Stokes photons thereby resulting in the
excitation of the QRs.

Figure~3(a) of the main text \cite{main} %\ref{Fig-QR-three-pulses} 
illustrates pump and Stokes radio-wave pulses
(red and purple arrows) propagating in the $x$-$y$ plane, and
the optical LG pulses propagating along the $z$ axis.
The blue colors denote the depth of the optical pulse potential.
The optical pulses lift the cylindrical symmetry of the trapping optical potential,
Eq.~(5) in \cite{main}, and allows quantum transitions with
$\Delta m_{\ell} = \pm 2 L$, i.e., it causes a rotational-kick to be applied to the QR.
Figure~3(b) in \cite{main} depicts quantum transitions between the ground state
with $m_{\ell} = 0$ and excited states with $m'_{\ell} = \pm 2 L$.
The stimulated Raman scattering process is modeled using a time-dependent Hamiltonian 
with matrix elements given by Eq.~(16) in the main text \cite{main}.

Consider Rabi oscillations resulting from the stimulated Raman process 
for the case when the low energy level is the $| m_{\ell} = 0 \rangle$ state, the 
high-energy levels are $| m_{\ell} = \pm 2L \rangle$ states, and
$$
  \epsilon_{2L} = \epsilon_{-2L},
  \quad
  \omega_{ps} = \frac{\epsilon_{2L} - \epsilon_0}{\hbar}.
$$
The temporal evolution of the QR wave function, starting with the initial wave function $|0\rangle$, is
specified by the time-dependent wave function
\begin{eqnarray}
  \big| \psi(t) \big\rangle =
  \cos\bigg(\frac{\Omega_R t}{2}\bigg)~
  \big| 0 \big\rangle -
  \frac{i}{\sqrt{2}}~
  \sin\bigg(\frac{\Omega_R t}{2}\bigg)
%  \nonumber \\ && \times
  \Big\{
      \big| 2L \big\rangle +
      \big| -2L \big\rangle
  \Big\},
  \label{eq:WF-Rabi}
\end{eqnarray}
where the Rabi frequency is
\begin{eqnarray}
  \Omega_R &=&
  \frac{2 \sqrt{2} {\mathcal{V}}}{\hbar}.
  \label{eq:Omega-Rabi}
\end{eqnarray}

\begin{figure}[htb]
%H=6.28, L=14.65
\centering
  \subfigure[]
  {\includegraphics[width=0.8\linewidth,angle=0]
   {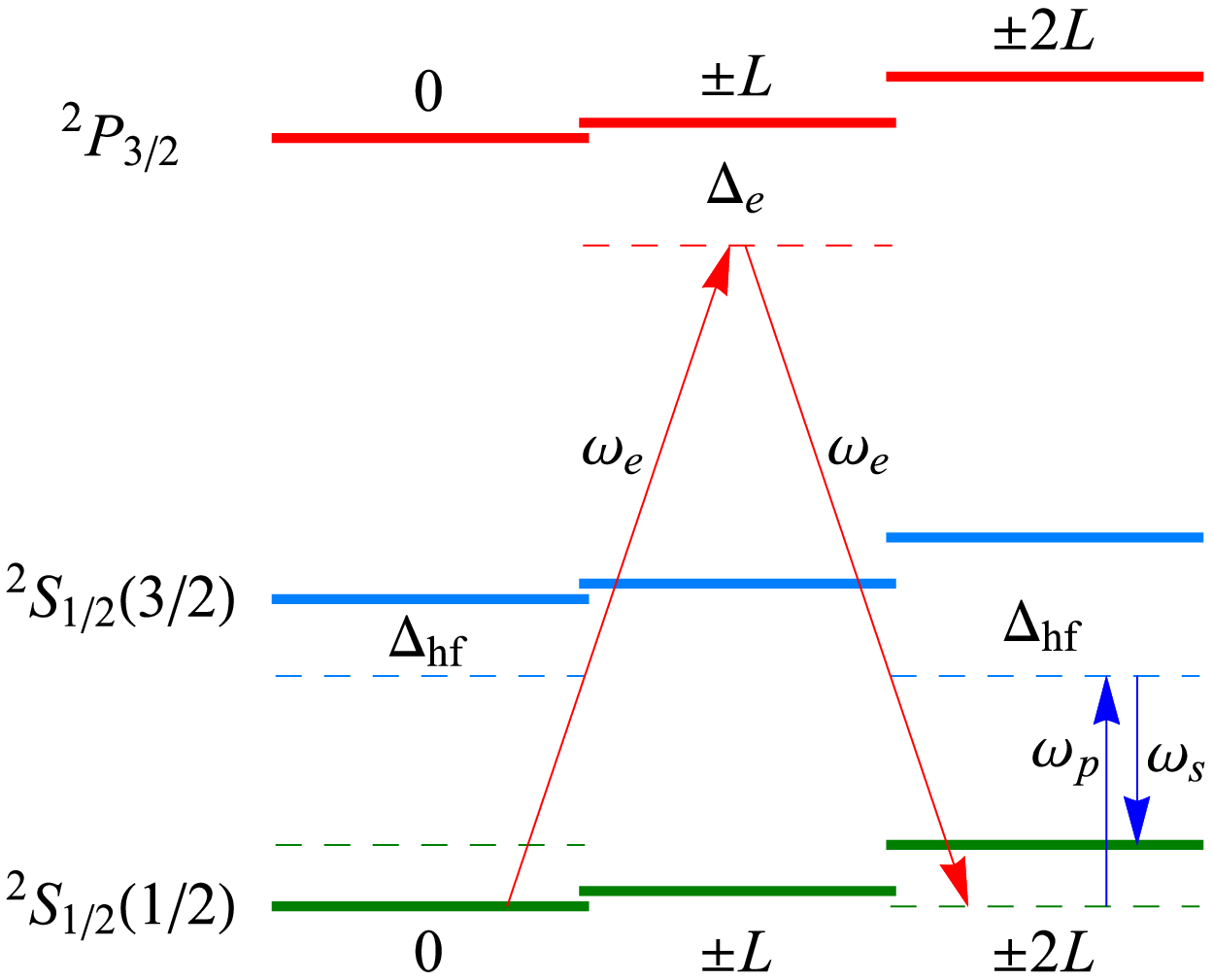}
   \label{Fig-diagram-1st}}
  \subfigure[]
  {\includegraphics[width=0.8\linewidth,angle=0]
   {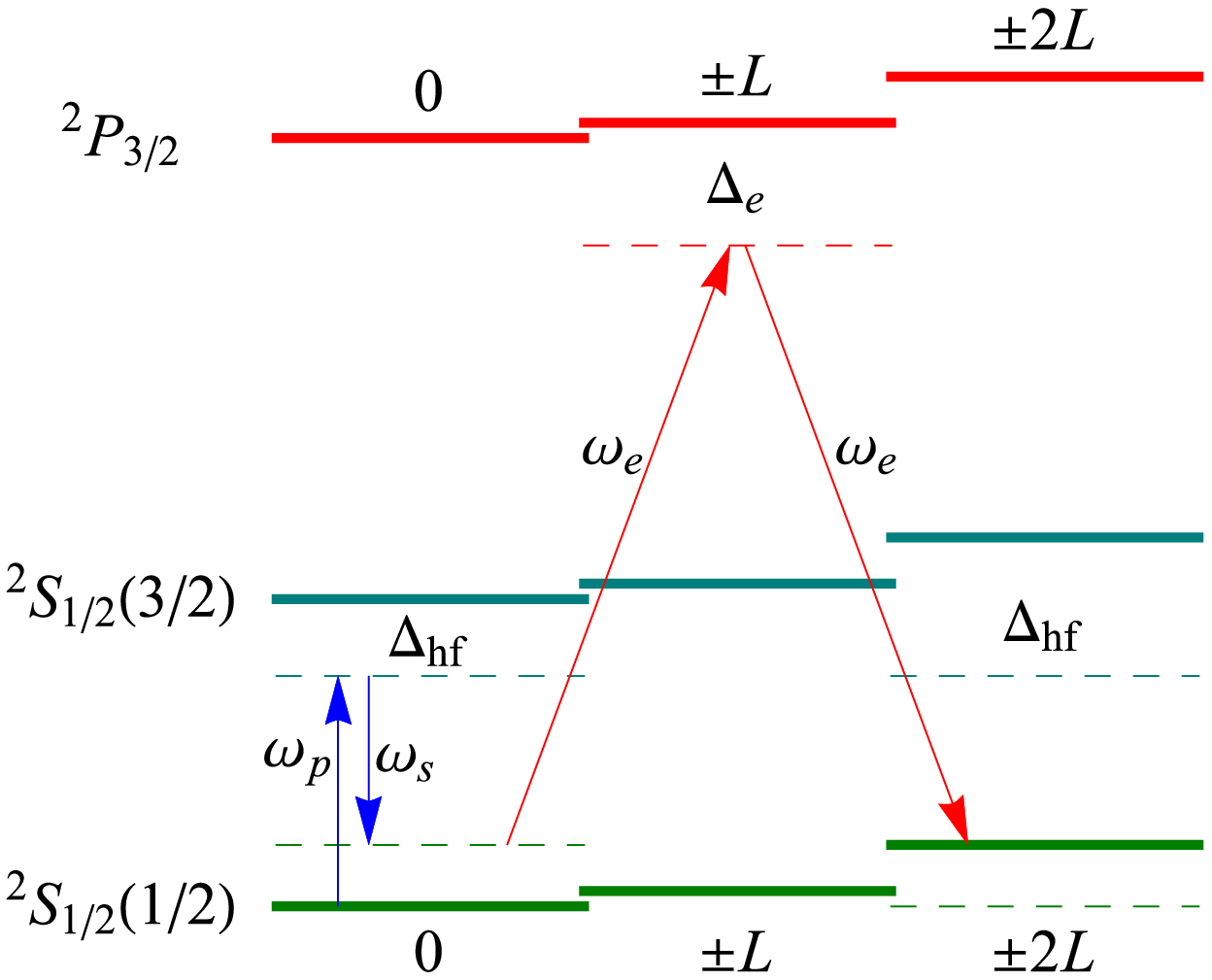}
   \label{Fig-diagram-2nd}}
 \caption{
   Far-off resonance stimulated Raman scattering due to pump and Stokes
   radiofrequency waves (blue arrows) and optical rotation-kicking optical pulses
   (red arrows). The two diagrams of processes that contribute are shown. 
   $\Delta_e$ is a detuning of $\omega_e$ from resonance
   with the ${}^{2}$S$_{1/2} \to {}^{2}$P$_{3/2}$ quantum transition, and
   $\Delta_{\mathrm{hf}}$ is a detuning of $\omega_p$ from resonance of the
   ${}^{2}$S$_{1/2}(F=1/2) \to {}^{2}$S$_{1/2}(F=3/2)$ quantum transition.
   The detuning $\Delta_e$ and $\Delta_{\mathrm{hf}}$, and
   the frequency of the quantum transition
   $\omega_{2L,0} = (\epsilon_{2 L} - \epsilon_0)/\hbar$
   satisfy the inequalities (\ref{inequality3:epsilon2L<<Dhf<<De}),
   therefore the optical pulses are taken to have the same detuning
   $\Delta_e$ in both diagrams (a) and (b).}
 \label{Fig-diagrams-1st-2nd}
\end{figure}

Consider an alkali atoms in the ground ${}^{2}$S$_{1/2}$ state trapped
in the Laguerre-Gaussian trapping potential (7) in the main text \cite{main}.
The atoms are illuminated by pump and Stokes radiofrequency waves far 
detuned by a detuning $\Delta_{\mathrm{hf}}$ from the hyperfine splitting, and
a laser beam far detuned from the excitation frequency of the ${}^{2}{\text{P}}_{3/2}$
state by a detuning $\Delta_e$.  Diagrams contributing to the quantum transitions 
$|0\rangle \to | \pm 2L \rangle$ in fourth-order perturbation theory are illustrated 
in Figs.~\ref{Fig-diagram-1st} and \ref{Fig-diagram-2nd}.
The frequencies of the pump and Stokes pulses are
$\omega_p$ and $\omega_s$, and the frequency of the laser pulse
is $\omega_e$.  The detuning frequencies from the resonances are given by
\begin{eqnarray}
  \Delta_e = \omega_e - \omega\big({}^{2}{\text{P}}_{3/2} - {}^{2}{\text{S}}_{1/2}\big),
  \ \ \
  \Delta_{\mathrm{hf}} = \omega_p - \omega_{\mathrm{hf}}
  \gtrapprox \omega_s - \omega_{\mathrm{hf}},
  \label{eq:Delta_e-Delta_hf}
\end{eqnarray}
where $\omega\big({}^{2}{\text{P}}_{3/2} - {}^{2}{\text{S}}_{1/2}\big)$ is a resonant
frequency of the ${}^{2}$S$_{1/2}$ $\to$ ${}^{2}$P$_{3/2}$
quantum transition, and $\omega_{\mathrm{hf}}$ is
the hyperfine splitting.  We assume that
\begin{eqnarray}
  \big| \Delta_e \big| ~\gg~
  \big| \Delta_{\mathrm{hf}} \big| ~\gg~
  \omega_{2L,0},
  \label{inequality3:epsilon2L<<Dhf<<De}
\end{eqnarray}
where $\omega_{2L,0}$ is the frequency of the quantum transition
$| 0 \rangle \to | 2L \rangle$. Hence, we can neglect $\omega_{2L,0}$ compared with
$\Delta_e$, and take the same detuning $\Delta_e$ of $\omega_e$ from resonance
with the ${}^{2}$S$_{1/2}$ $\to$ ${}^{2}$P$_{3/2}$ transition [as shown in 
Figs.~\ref{Fig-diagram-1st} and \ref{Fig-diagram-2nd}].  Moreover, we assume that the detuning
$\omega_p - \omega_{\mathrm{hf}}$ of the pump pulse is approximately equal to the detuning
$\omega_s - \omega_{\mathrm{hf}}$ of the Stokes pulse.

Interactions of the trapped atom with the pump, Stokes
and optical pulses are described by the Hamiltonians,
\begin{eqnarray}
  H_{\mu} = -g \mu_B \mbfs \cdot \mbfB_{\mu}(t),
  \ \ \ \ \
  H_e = -\mbfp \cdot \mbfE_e(\mbfr,t),
  \label{eq:Hp-Hs-He}
\end{eqnarray}
where $\mu = p,s$ for the pump and Stokes pulses,
$\mbfs$ is the spin-$1/2$ vector operator,
and $\mbfp$ is the atomic electric dipole operator.
The subscript $e$ here and in Eq.~(17) of \cite{main} symbolizes the electric dipole
interaction of the atom with the optical pulse, and the subscript $b$ symbolizes the
magnetic dipole interaction of the atoms with the pump and Stokes
radio-frequency pulses.
The electric field $\mbfE_e(\mbfr,t)$ and the magnetic field
$\mbfB_{\mu}(t)$ are given by
\begin{eqnarray*}
  &&
  \mbfB_{\mu}(t) = \mbfB_{\mu}^{(0)} \cos(\omega_{\mu} t),
  \\
  &&
  \mbfE_e(\mbfr,t) =
  \frac{1}{2}
  \big\{
      u_{L,0}(\mbfr) +
      u_{-L,0}(\mbfr)
  \big\}
  \cos(k_e z)
  e^{-i \omega_e t}
  \mbfe_x +
  {\mathrm{c.c.}},
\end{eqnarray*}
where $u_{L,0}(\mbfr)$ is given by Eq.~(2) in the main text \cite{main}.

%---------------- 4th order diagram ---------------------------

When a trapped atom absorbs and reemit an optical pulse
photon, it gets a rotational-kicking quantum transition with
$\Delta m_{\ell} = \pm 2 L$, where $m_{\ell}$ is the orbital
quantum number of orbital motion of the atom around
the $z$ axis, and $L$ is the orbital moment of the optical
Laguerre-Gaussian (LG) pulse, see the main text \cite{main}
for details. From the other side, absorbing a pump photon
and reemitting a Stokes photon, the atom gets the energy
needed for quantum transitions. In this section, we apply
fourth order perturbation theory and derive an effective
Hamiltonian (16) in the main text \cite{main}.
This derivation is rather standard, but cumbersome.

We derive here an effective Hamiltonian describing Rabi
oscillations for the QR in the $m_{\ell} = 0$ and $\pm 2L$ states.
For this purpose, we assume that the low-energy states of
the trapped atom are $|{}^{2}{\text{S}}_{1/2}(1/2),m_{\ell},m_F\rangle$
states with the electronic configuration $^2$S$_{1/2}$($F = 1/2$),
magnetic quantum number $m_F = \pm 1/2$ and the motion of
the center of mass parametrized by $m_{\ell} = 0, \pm 2L$.
High-energy states are $|{}^{2}{\text{S}}_{1/2}(3/2),m_{\ell},m_F\rangle$
with electronic configuration $^2$S$_{1/2}$($F = 3/2$),
$m_{\ell} = 0, \pm 2L$ and $m_F = \pm 1/2$,
and $|{}^{2}{\text{P}}_{3/2},m_{\ell},m_F\rangle$ with
electronic configuration $^2$P$_{3/2}$, $m_{\ell} = \pm L$
and $m_F = \pm 1/2$, as illustrated in Fig.~\ref{Fig-diagrams-1st-2nd}.
In this section, we use the following basis,
\begin{subequations}
\begin{eqnarray}
  |0\rangle &=&
  \big|
      {}^{2}{\text{S}}_{1/2}
      \big(1/2\big),~
      0,~
      m_F
  \big\rangle,
  \label{eq:WF-1}
  \\
  |1\rangle &=&
  \frac{1}{\sqrt{2}}~
  \Big\{
  \big|
      {}^{2}{\text{S}}_{1/2}
      \big(1/2\big),~
      2L,~
      m_F
  \big\rangle +
  \nonumber \\ &&
  \big|
      {}^{2}{\text{S}}_{1/2}
      \big(1/2\big),~
      -2L,~
      m_F
  \big\rangle
  \Big\},
  \label{eq:WF-2}
  \\
  |2\rangle &=&
  e^{i \omega_p t}
  \big|
      {}^{2}{\text{S}}_{1/2}
      \big(3/2\big),~
      0,~
      m_F
  \big\rangle,
  \label{eq:WF-3}
  \\
  | 3 \rangle &=&
  \frac{e^{i \omega_p t}}{\sqrt{2}}
  \Big\{
  \big|
      {}^{2}{\text{S}}_{1/2}
      \big(3/2\big),~
      2L,~
      m_F
  \big\rangle +
  \nonumber \\ &&
  \big|
      {}^{2}{\text{S}}_{1/2}
      \big(3/2\big),~
      -2L,~
      m_F
  \big\rangle
  \Big\},
  \label{eq:WF-4}
  \\
  | 4 \rangle &=&
  \frac{e^{i \omega_e t}}{\sqrt{2}}
  \Big\{
  \big|
      {}^{2}{\text{P}}_{3/2},~
      L,~
      m_F
  \big\rangle +
  \big|
      {}^{2}{\text{P}}_{3/2},~
      -L,~
      m_F
  \big\rangle
  \Big\}.
  \label{eq:WF-5}
\end{eqnarray}
  \label{subeqs:WF-12345}
\end{subequations}
The ``non-perturbed'' Hamiltonian of the trapped atom
without the pump, Stokes and optical pulses is given by
the matrix elements,
\begin{eqnarray}
  \langle \nu | H_0 \nu' \rangle &=&
  \veps_{\nu} \delta_{\nu,\nu'},
  \label{eq:H0}
\end{eqnarray}
where
\begin{eqnarray*}
  \veps_0 = 0,
  \ \
  \veps_1 = \epsilon_{2L},
  \ \
  \veps_2 = \hbar \Delta_{\mathrm{hf}},
  \ \
  \veps_3 = \hbar \Delta_{\mathrm{hf}} + \epsilon_{2L},
  \ \
  \epsilon_4 = \hbar \Delta_e.
\end{eqnarray*}
The energy of the rotational motion of the atom is
$\epsilon_{m_{\ell}} = m_{\ell}^{2} C(r_l)$, where
$C(r_l) = \hbar^2/(2 M r_{l}^{2})$ and $r_l$ is
the radius of the classical circular trajectory.
$\Delta_{\mathrm{hf}} = \omega_p - \epsilon_{\mathrm{hf}}/\hbar$ and
$\Delta_e = \omega_e - \epsilon_e/\hbar$ are
the detuning of the pump and optical frequencies from
the resonant frequencies of the quantum transitions
$|0\rangle \to |3\rangle$ and $|0\rangle \to |4\rangle$,
respectively.

Nontrivial matrix elements of $H_p$, $H_s$ and $H_e$,
Eq.~(\ref{eq:Hp-Hs-He}), are
\begin{subequations}
\begin{eqnarray}
  \big\langle
      1
  \big|
      H_{\mu}
  \big|
      3
  \big\rangle
  &=&
  \big\langle
      2
  \big|
      H_{\mu}
  \big|
      4
  \big\rangle
  =
  g \mu_B B_{\mu}
  \big\langle
      1/2,m_F
  \big|
      s_x
  \big|
      3/2,m_F
  \big\rangle
  \nonumber \\ &=&
  \frac{\sqrt{2}}{3}~
  g \mu_B B_{\mu} e^{i (\omega_{\mu} - \omega_p) t} m_F,
  \label{eq:H-ps-ME}
  \\
  \big\langle
      1
  \big|
      H_e
  \big|
      5
  \big\rangle
  &=&
  \big\langle
      2
  \big|
      H_e
  \big|
      5
  \big\rangle
  =
  \frac{1}{2}~
  \big[
      u_{L,0}(r_l,0,0) + u_{-L,0}(r_l,0,0)
  \big]~
  \nonumber \\ && \times
  \big\langle
      {}^{2}{\text{S}}_{1/2},m_F
  \big|
      p_x
  \big|
      {}^{2}{\text{P}}_{3/2},m_F
  \big\rangle,
  \label{eq:H-e-ME}
\end{eqnarray}
\end{subequations}
where $u_{L,0}(r,\phi,z)$ are given by Eq.~(2) in
the main text \cite{main}.
Here we chose the $x$ axis as a quantization axis.

%%%%%%%%%%%%%%%%%%%%
\subsection{Adiabatic elimination of the $^2$S$_{1/2}$($F = 3/2)$
  hyperfine state}

As a first step, we apply the following unitary transformations,
\begin{eqnarray*}
  | \psi_0 \rangle = u_b |0\rangle - v_b |2\rangle,
  \ \ \ \ \
  | \psi_1 \rangle = u_b |1\rangle - v_b |3\rangle,
  \\
  | \psi_2 \rangle = v_{b}^{*} |0\rangle + u_b |2\rangle,
  \ \ \ \ \
  | \psi_3 \rangle = v_{b}^{*} |1\rangle + v_b |3\rangle,
\end{eqnarray*}
and $|\psi_4\rangle = |4\rangle$, where
\begin{eqnarray}
  v_b = \frac{\sqrt{2}}{3} ~ \frac{g \mu_B \big(B_p + B_s e^{-i \omega_{ps}}\big)}{\hbar \Delta_{\mathrm{hf}}},
  \ \ \ \ \
  u_b = \sqrt{1 - |v_b|^{2}},
  \label{eq:vb-ub-def}
\end{eqnarray}
and $\omega_{ps} = \omega_p - \omega_s$.
We assume here that $|v_b| \ll 1$ and keep terms up to $v_{b}^{2}$, and neglect $v_{b}^{n}$ with $n \geq 3$.
This transformation makes $H_0+H_p+H_s$ diagonal,
\begin{eqnarray*}
  \langle \psi_n | H_0 + H_p + H_s | \psi_{n'} \rangle &=&
  \tilde\veps_n
  \delta_{n,n'},
\end{eqnarray*}
where
\begin{eqnarray*}
  \tilde\veps_0 &=&
  -\Delta_{\mathrm{hf}} |v_b|^{2},
  \\
  \tilde\veps_1 &=&
  \epsilon_{2L} - \Delta_{\mathrm{hf}} |v_b|^{2},
  \\
  \tilde\veps_2 &=&
  \Delta_{\mathrm{hf}} \big( 1 + |v_b|^{2} \big),
  \\
  \tilde\veps_3 &=&
  \epsilon_{2L} + \Delta_{\mathrm{hf}} \big( 1 + |v_b|^{2} \big),
  \\
  \tilde\veps_4 &=&
  \tilde\veps_4.
\end{eqnarray*}
Matrix elements of $H_e$ are
\begin{eqnarray}
  h_e &\equiv&
  \langle \psi_0 | H_e | \psi_4 \rangle =
  \langle \psi_1 | H_e | \psi_4 \rangle
  \nonumber \\ &=&
  \frac{1}{2}~
  \big[
      u_{L,0}(r_l,0,0) + u_{-L,0}(r_l,0,0)
  \big]~
  \times \nonumber \\ &&
  \big\langle
      {}^{2}{\text{S}}_{1/2},m_F
  \big|
      p_x
  \big|
      {}^{2}{\text{P}}_{3/2},m_F
  \big\rangle~
  \bigg( 1 - \frac{|v_b|^{2}}{2} \bigg),
  \label{eq:he-def}
\end{eqnarray}
where $u_{L,0}(r,\phi,z)$ is given by Eq.~(2) in the main text \cite{main}.

%%%%%%%%%%%%%%%%%%%%
\subsection{Adiabatic elimination of the $^2$P$_{3/2}$
  excited state}

In a second step, we apply the following unitary transformations,
\begin{eqnarray*}
  \big| \tilde\psi_0 \big\rangle &=&
  u_{g} \big| \psi_0 \big\rangle -
  \frac{v_{e}^{2}}{2}~ \big| \psi_1 \big\rangle -
  v_e \big| \psi_4 \big\rangle,
  \\
  \big| \tilde\psi_1 \big\rangle &=&
  -\frac{v_{e}^{2}}{2}~ \big| \psi_0 \big\rangle +
  u_{g} \big| \psi_1 \big\rangle -
  v_e \big| \psi_4 \big\rangle,
  \\
  \big| \tilde\psi_4 \big\rangle &=&
  v_e \big| \psi_0 \big\rangle +
  v_e \big| \psi_1 \big\rangle +
  u_e \big| \psi_4 \big\rangle,
\end{eqnarray*}
where
\begin{eqnarray*}
  v_e = \frac{h_e}{\Delta_e},
  \ \ \ \ \
  u_g = \sqrt{1 - |v_e|^{2}},
  \ \ \ \ \
  u_e = \sqrt{1- 2 |v_e|^{2}}.
\end{eqnarray*}
We assume here that $|v_e| \ll 1$ and keep terms up to $v_{e}^{2}$
and neglect terms proportional to $v_{e}^{n}$ with $n \geq 3$.
Then the transformed Hamiltonian $H = H_0 + H_p + H_s + H_e$
is given by the matrix elements,
\begin{eqnarray*}
  \big\langle \tilde\psi_{n} \big|
      H
  \big| \tilde\psi_{n'} \big\rangle
  &=&
  \tilde\veps_n \delta_{n,n'} -
  \frac{2 |h_e|^{2}}{\Delta_e},
  \\
  \big\langle \tilde\psi_4 \big|
      H
  \big| \tilde\psi_4 \big\rangle
  &=&
  \Delta_e +
  \frac{4 |h_e|^{2}}{\Delta_e},
  \\
  \big\langle \tilde\psi_n \big|
      H
  \big| \tilde\psi_4 \big\rangle
  &=& 0,
\end{eqnarray*}
where $n,n' = 0, 1$.

Omitting the high-energy state $|\tilde\psi_4\rangle$,
we get the two-level effective Hamiltonian describing
Rabi oscillations,
\begin{eqnarray}
  H =
  \left(
    \begin{array}{cc}
      0
      &
      - 2 |h_e|^{2}/\Delta_e
      \\
      - 2 |h_e|^{2}/\Delta_e
      &
      \epsilon_{2L}
    \end{array}
  \right),
  \label{eq:H-Rabi-2}
\end{eqnarray}
where we shift the chemical potential by adding the term
$$
  \bigg(
       \Delta_{\mathrm{hf}} |v_b|^{2} +
       \frac{2 |h_e|^{2}}{\Delta_e}
  \bigg)
  \left(
    \begin{array}{cc}
      1 & 0
      \\
      0 & 1
    \end{array}
  \right).
$$

Taking into account Eqs.~(\ref{eq:he-def}) and (\ref{eq:vb-ub-def}),
we get
\begin{eqnarray}
  &&
  \frac{2 |h_e|^{2}}{\Delta_e} =
  \frac{\alpha(\omega_e)}{2}~
  \big|
      u_{L,0}(r_l,0,0) + u_{-L,0}(r_l,0,0)
  \big|^{2}
  \times
  \nonumber \\ && ~~~~~
  \Bigg\{
        1 -
        \frac{1}{3}~\frac{g^2 \mu_{B}^{2}}{\hbar^2 \Delta_{\mathrm{hf}}^{2}}~
        \Big[
           B_{p}^{2} + B_{s}^{2} +
           2 B_p B_s \cos(\omega_{ps} t)
       \Big]
  \Bigg\}.
  \label{eq:effective-interaction}
\end{eqnarray}
On the right hand side of Eq.~(\ref{eq:effective-interaction}) there are
time-independent and time-dependent terms. The former can be considered
as weak perturbation which does not contribute to the Rabi oscillations,
whereas the second term gives rise to Rabi oscillations.
Off-diagonal part of the Hamiltonian (\ref{eq:H-Rabi-2}) after omitting
the time-independent terms gives us the Rabi oscillation Hamiltonian
(16) in the main text \cite{main}.

%%%%%%%%%%%%%%%%%%%%%%%%%%%%%%
\section{Rotation Sensor Accuracy}  \label{sec:accuracy}

\begin{figure}%[htb]
%H=6.28, L=14.65
\centering
  \includegraphics[width = 60 mm,angle=0]
   {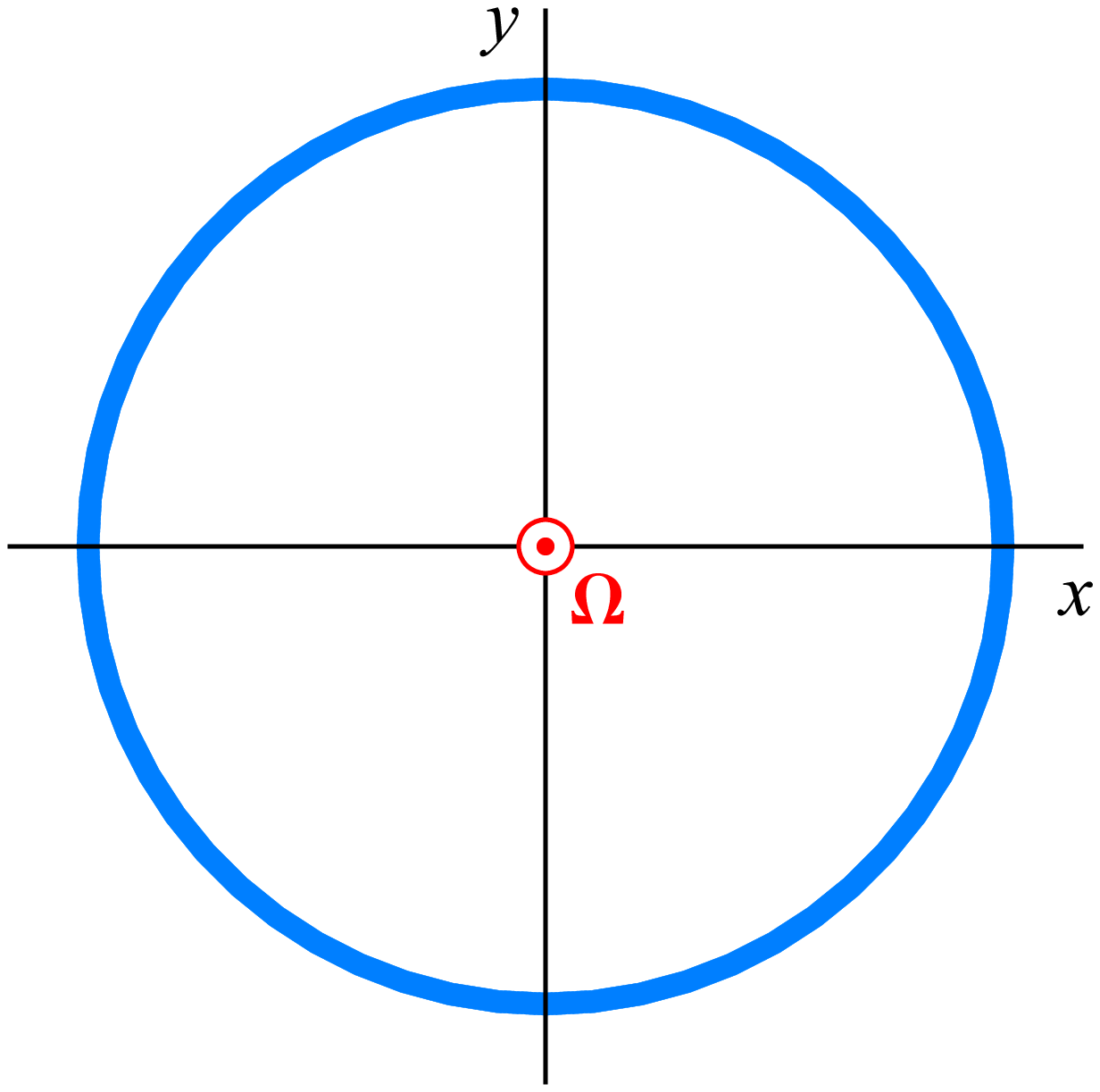}
\caption{QR (blue ellipse) subject to a rotation with angular
  velocity $\boldsymbol\Omega = \Omega {\bf e}_z$.}
\label{Fig-rotation}
\end{figure}

When QRs are placed in a non-inertial frame rotating with angular
velocity $\boldsymbol\Omega = \Omega \mbfe_z$ (as described
in the main text \cite{main} and as illustrated in Fig.~\ref{Fig-rotation}),
they can be used as a high accuracy rotation sensor to determine $\Omega$.
The accuracy of the rotation sensor due to uncertainty of
the pump and Stokes frequencies is analyzed in
the main text. Here we derive the uncertainty in the angular 
velocity due to Rabi frequency fluctuations and
due to shot noise in the pump and Stokes pulses.

%---------------- acceleration rotation ---------------------------

%%%%%%%%%%%%%%%%%%%%
\subsection{Uncertainty due to Rabi frequency fluctuation}

The energies $\epsilon_{m_{\ell}} = m_{\ell}^{2} C(r_l)$ do
not depend on the laser frequency and intensity.
Therefore, the frequencies of the quantum transitions
$| m_{\ell} \rangle \to | m'_{\ell} \rangle$ are independent
of the laser frequency and intensity.
Hence, the only source of uncertainty is fluctuations in
frequencies of the Stokes, pump and ``kick'' pulses
which results in fluctuations in the Rabi oscillations.
Indeed, fluctuations of frequencies result in fluctuations
in the Rabi frequency (\ref{eq:Omega-Rabi}),
\begin{eqnarray}
  \phi_R =
  \Omega_R \tau_{\pi} =
  \pi \pm \delta\phi_{\omega},
  \label{eq:phi-R-frequency}
\end{eqnarray}
where
\begin{eqnarray}
  \delta\phi_{\omega} &=&
  \sqrt{\sum_{\kappa = p,s}
        \bigg(
             \frac{\partial \phi_R}{\partial \omega_{\kappa}}~
             \delta\omega_{\kappa}
        \bigg)^{2}}
  \nonumber \\  &=&
  \phi_R
  \sqrt{\bigg( \frac{\delta \omega_p}{\Delta_{\mathrm{hf}}} \bigg)^{2} +
        \bigg( \frac{\delta \omega_s}{\Delta_{\mathrm{hf}}} \bigg)^{2}}.
  \label{eq:delta-phi-omega}
\end{eqnarray}
Taking $\delta \omega_p \approx \delta \omega_s = 2.86 \times 10^{-9}$~s$^{-1}$
(see the main text \cite{main}) and $\Delta_{\mathrm{hf}} = 1.26 \times 10^8$~s$^{-1}$,
we get
\begin{eqnarray}
  \frac{\delta \phi_{\omega}}{\phi_R} &=&
  3.21 \times 10^{-17}.
  \label{eq:delta-phi-omega-res}
\end{eqnarray}

When the energy levels $\epsilon_{\pm2L}$ are not degenerate,
but $\Delta \epsilon = \epsilon_{2L} - \epsilon_{-2L}$ is small
in comparison to $\epsilon_{2L} - \epsilon_0$,
the difference $\Delta\phi_{\epsilon} \equiv \phi_{2L} - \phi_{-2L}$ is
\begin{eqnarray}
  \Delta\phi_{\epsilon} &=&
  \frac{\Delta \epsilon}{4 \hbar \Omega_R},
  \label{eq:delta-phi-epsilon}
\end{eqnarray}
where $\Omega_R$ is the Rabi frequency (\ref{eq:Omega-Rabi}).
The levels $\epsilon_{\pm2L}$ cam be distinguished when
$\Delta \phi_{\epsilon} > \delta \phi_{\omega}$, or
\begin{eqnarray}
  \frac{\Delta \epsilon}{C(r_l)} >
  \frac{\delta \epsilon_{\omega}}{C(r_l)} =
  \frac{4 \hbar \Omega_R}{C(r_l)}~
  \delta \phi_{\omega} =
  5.998 \times 10^{-17},
  \label{eq:uncertainty-splitting-omega}
\end{eqnarray}
where we take $\Omega_R = 3.142$~s$^{-1}$.
Then, taking $C(r_l) = k_B \times 0.1613$~nK
(which corresponds to the Laguerre-Gaussian beam with $l = 5$,
$p = 0$ and $w_0 = 10~\mu$m, see Eq.~(2) in the main text \cite{main}),
we get the uncertainty of the energy splitting
$\delta\epsilon_{\omega}$ as
\begin{equation}   \label{eq:uncertainty-Omega-frequency}
  \delta \epsilon_{\omega} = \hbar  \times 1.267 \times 10^{-15}~{\text{s}}^{-1}.
\end{equation}

Knowing the uncertainty $\delta \epsilon_{\omega}$ of
the energy splitting, we are able to estimate the uncertainty
$\delta \Omega_{\omega}$ of the rotation sensor,
\begin{eqnarray}
  \delta \Omega_{\omega} &=&
  \frac{1}{\sqrt{N}} ~ \frac{\delta \epsilon_{\omega}}{4 L \hbar},
  \label{eq:delta-Omega-omega}
\end{eqnarray}
where $N = 2 j_{\max} + 1$ is the number of singly-occupied rings filled with QR atoms.
When $L = 25$ and $N = 161$, then
\begin{eqnarray*}
  \delta \Omega_{\omega} &=&
  9.985 \times 10^{-19}~{\text{s}}^{-1}.
\end{eqnarray*}
Comparing this result with $\delta \Omega$ in Eq.~(23) in the main text \cite{main},
one can see that uncertainty in $\Omega$ due to
fluctuations of the Rabi frequency is very small with respect
to the uncertainty $\delta \Omega$ due to the fluctuations of the pump and
Stokes frequencies.

\bigskip
%%%%%%%%%%%%%%%%%%%%
\subsection{Uncertainty due to shot noise in the pump and Stokes pulses}

Another source of uncertainty arises from shot noise in
the Stokes, pump and kick pulses. Shot noise results in
fluctuations in the position and amplitude of the population
oscillations of the Ramsey fringes because of fluctuation of
the Rabi frequencies,
\begin{eqnarray}
  \phi_R &=&
  \Omega_R \tau_{\pi} =
  \pi \pm \delta \phi_I,
  \label{eq:phi-R-shot-noise}
\end{eqnarray}
where
\begin{eqnarray}
  \delta \phi_I &\approx&
  \pi \,
  \bigg(
       \frac{1}{\sqrt{N_p}} +
       \frac{1}{\sqrt{N_s}}
  \bigg),
  \label{eq:delta-phi-shot-noise}
\end{eqnarray}
where $N_p$ and $N_s$ are the number of photons
in the pump and Stokes pulses during the time
$\tau_{\pi} = \pi/\Omega_R$.
When $N_p \sim N_s \sim 10^{29}$,
then
$$
  \delta \phi_I = 1.987 \times 10^{-14}.
$$
Uncertainty in the energy splitting can be estimated in a fashion similar to
Eq.~(\ref{eq:uncertainty-splitting-omega}),
\begin{eqnarray}
  \frac{\Delta \epsilon}{4 \hbar \Omega_R} >
  \frac{\delta \epsilon_{I}}{4 \hbar \Omega_R} =
  \frac{\delta \phi_I}{\phi_R} =
  6.325 \times 10^{-15}.
  \label{eq:uncertainty-splitting-shot-noise}
\end{eqnarray}
For $\Omega_R = 3.142$~s$^{-1}$, we get
$\delta\epsilon_I = \hbar \times 7.949 \times 10^{-14}$~s$^{-1}$.

Knowing the uncertainty $\delta \epsilon_{I}$ of
the energy splitting, we are able to estimate the uncertainty
$\delta \Omega_{I}$ of the rotation sensor,
\begin{eqnarray}
  \delta \Omega_I = \frac{1}{\sqrt{N}} ~
  \frac{\delta \epsilon_I}{4 L \hbar},
  \label{eq:delta-Omega-shot-noise}
\end{eqnarray}
where  $N = 2 j_{\max} + 1$ is the number of singly-occupied rings filled with QR atoms.
When $L = 25$ and $N = 161$, the uncertainty is
\begin{eqnarray*}
  \delta \Omega_I &=&
  6.265 \times 10^{-17}~{\text{s}}^{-1}.
\end{eqnarray*}
Comparing these results with Eq.~(23) in the main text \cite{main},
one can see that the uncertainty in $\Omega$ due to
shot noise in the pump, Stokes and kick pulses is of the same order
of magnitude as the uncertainties due to fluctuations the pump and Stokes
frequencies.

%---------------- QR-incline ---------------------------
\begin{figure}[htb]
%H=6.28, L=14.65
\centering
  \includegraphics[width = 70 mm,angle=0]
   {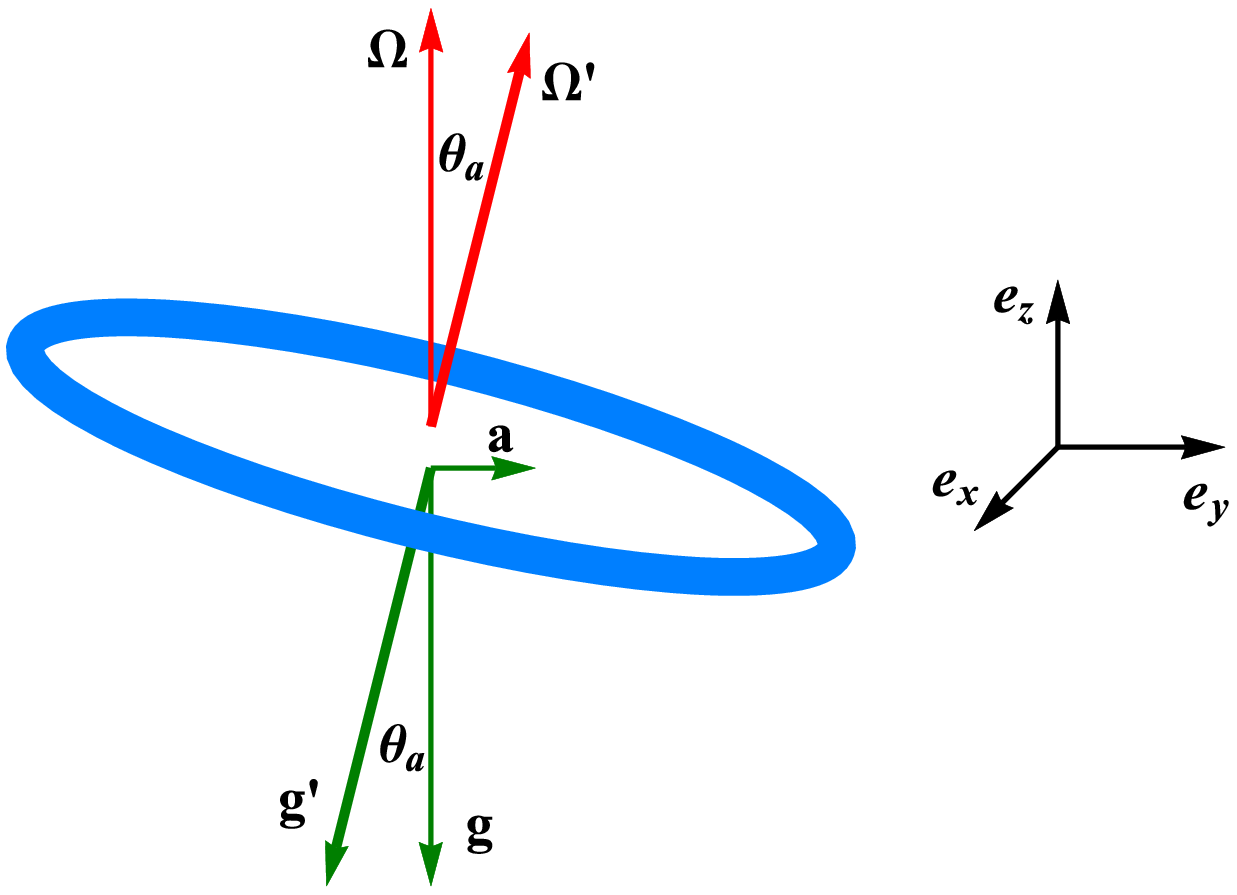}
 \caption{
   Elimination of the in-plane acceleration $\mbfa$ by inclining
   the QRs. Here ${\mathbf{g}} = -g \mbfe_z$ is the acceleration
   due to gravity, $\boldsymbol\Omega$ is the angular velocity,
   and ${\mathbf{g}}' = {\mathbf{g}} - \mbfa$ is the total acceleration,
   where $\mbfe_x$, $\mbfe_y$ and $\mbfe_z$ are unit vectors
   parallel to the $x$-, $y$- and $z$-axes.
   $\boldsymbol\Omega' = \Omega' \mbfe'_z$, where
   $\Omega' = (\boldsymbol\Omega \cdot \mbfe'_z)$ and
   the unit vector $\mbfe'_z$ is antiparallel to ${\mathbf{g}}'$.
   The angle between ${\mathbf{g}}$ and ${\mathbf{g}}'$
   is $\theta_a$. Blue ellipse is the QR placed in the plane 
   perpendicular to ${\mathbf{g}}'$.}
\label{Fig-QR-incline}
\end{figure}

%%%%%%%%%%%%%%%%%%%%%%%%%%%%%%
\section{Discriminating against in-plane acceleration}  \label{sec:acceleration}

An additional term $H_g = -M \mathbf{g} \cdot \mbfr$ must be added into the QR Hamiltonian
to model the affects of a gravitational field $\mathbf{g}$ \cite{Kuzmenko_19}. This preserves 
the rotational symmetry in the plane perpendicular to $\mathbf{g}$ (the $x$-$y$ plane
in Fig.~\ref{Fig-QR-incline}),
but lifts the rotational symmetry in the $x$-$z$ and $y$-$z$ planes.
As a result, the QRs rotating in the $x$-$y$ plane clockwise and
counterclockwise with the same quantum number $|m_{\ell}|$
have the same energy.
When the QRs are placed in the $x$-$z$ or $y$-$z$ plane
(such that $\mathbf{g}$ is in the plane of the QRs), the degeneracy
of the quantum states $|m_{\ell}\rangle$ and $|-m_{\ell}\rangle$
are split and the splitting depends on $m_{\ell}$.
Hence, placing the QRs in the $x$-$y$ plane, we obtain the energy
splitting of the levels caused just by $\Omega_z$ (the $z$ component
of the angular velocity $\boldsymbol\Omega$).

Now consider QRs placed in a non-inertial frame moving with
acceleration $\mbfa$.
The effective gravity $\mathbf{g}'$ in the non-inertial frame is given
by ${\mathbf{g}}' = {\mathbf{g}} - \mbfa$.
Therefore, when we turn the QRs to the plane perpendicular to
${\mathbf{g}}'$, we obtain the splitting of the energy levels caused
just by $\Omega' = \boldsymbol\Omega \cdot \mbfe'_z$, as illustrated
in Fig.~\ref{Fig-QR-incline}, where $\mbfe'_z$ is the unit vector along
$-\mathbf{g}'$.